\documentclass[12pt]{cernrep}
\usepackage{graphicx}
\begin{document}

\begin{minipage}{14cm}
\vskip 2cm
\hspace*{10.2cm} {\bf Preprint JINR}\\
\hspace*{10.5cm}{\bf E2-2004-178}\\
\hspace*{10.5cm}{\bf Dubna, 2004}\\
%\hspace*{10.1cm}{\bf hep-ph/0411229}\\

\end{minipage}
\vskip 4cm

\begin{center}
{\bf
 $Z$-SCALING AT RHIC}
% TITLE OF THE ARTICLE (IN CAPITAL LETTERS)}
\vskip 5mm
 M.V. Tokarev

\vskip 5mm

{\small
 {\it
Veksler and Baldin Laboratory of High Energies,\\
Joint Institute for Nuclear Research,\\
141980, Dubna, Moscow region, Russia}
\\
 {\it
E-mail: tokarev@sunhe.jinr.ru
}
}
\end{center}

\vskip 5mm

\begin{center}
\begin{minipage}{150mm}
\centerline{\bf Abstract}
The concept of $z$-scaling reflecting the general regularities of high-$p_T$
particle production is reviewed. Properties of data $z$-presentation are discussed. New
data on high-$p_T$ particle spectra obtained at the RHIC  are
analyzed in the framework of $z$-presentation. It was shown that
these experimental data confirm $z$-scaling. Predictions of strange
particle spectra are considered to be useful for understanding
of strangeness origin in mesons and baryons
and search for new physics phenomena at the RHIC.
\end{minipage}
\end{center}

\vspace*{2cm}
\begin{center}
{\it
    Submitted to "Physics of Particles and Nuclei, Letters"
%    Submitted to "Physical Review C"
% Submitted to the XV International Seminar on High Energy Physics\\
% "Relativistic Nuclear Physics  and Quantum Chromodynamics",
%September 25-29, 2000, Dubna, Russia
}
\end{center}

%\vskip 1cm
\newpage

{\section{Introduction}}

Search for scaling regularities in high energy particle collisions is always
to be a subject of intense investigations \cite{Feynman}-\cite{Brodsky}.
Commissioning of the Relativistic Heavy Ion Collider (RHIC) at the
Brookhaven National Laboratory (BNL) gives new possibilities to perform
experimental investigations in a new physics domain. The RHIC is a next generation
of a proton-proton colliders after ISR designed to accelerate protons at center
of mass energy range $\sqrt s = 50-500$~GeV aimed to clarify origin of proton's
spin and discover a new state of nuclear matter, Quark Gluon Plasma.

High energy of colliding particles and high transverse momentum
of produced particles are most suitable for precise QCD test of production
processes with hard probes like high-$p_T$ hadrons, direct photons and jets.
Therefore, search for general regularities of high-$p_T$
single inclusive particle spectra of hadron--hadron and  hadron--nucleus
collisions are of interest to establish  complementary restrictions
for theory.

The universal phenomenological description ($z$-scaling) of high-$p_T$ particle
production  cross sections in inclusive reactions is developed
in \cite{Zscal,zppg}. The approach is based on properties of particle structure,
their constituent interaction and particle formation such as locality,
self-similarity and fractality.
The scaling function $\psi$ and scaling variable $z$ are expressed via
experimental quantities such as the inclusive cross section
$Ed^3\sigma/dp^3$ and the multiplicity density of charged particles
$dN/d\eta$. Data $z$-presentation is found to reveal symmetry
properties (energy and angular independence, A-and F-dependence, power law).
The properties  of $\psi$ at high $z$ are assumed  to be
relevant to the structure of space-time at small scales \cite{Nottale,Mandelbrot,Imr}.
The function  $\psi(z)$ is interpreted as the probability density
to produce a particle with a formation length $z$.

In the report we present the results of analysis
of new data on high-$p_T$ particle spectra obtained at the RHIC.
The obtained results are compared with other ones based
on the data obtained at lower collision energy $\sqrt s $.
The results are considered as a new confirmation of $z$-scaling at the RHIC.

\vskip 0.5cm
{\section{Z-scaling }}

The idea of $z$-scaling is based on the assumptions \cite{Stavinsky}
that gross feature of inclusive particle distribution of the
process (\ref{eq:r1}) at high energies can be described in terms
of the corresponding kinematic characteristics
\begin{equation}
M_{1}+M_{2} \rightarrow m_1 + X
\label{eq:r1}
\end{equation}
of the constituent subprocess  written in the symbolic form (\ref{eq:r2})
\begin{equation}
(x_{1}M_{1}) + (x_{2}M_{2}) \rightarrow m_{1} +
(x_{1}M_{1}+x_{2}M_{2} + m_{2})
\label{eq:r2}
\end{equation}
satisfying the condition
\begin{equation}
(x_1P_1+x_2P_2-p)^2 =(x_1M_1+x_2M_2+m_2)^2.
\label{eq:r3}
\end{equation}
The equation is the expression of locality of  hadron interaction at
constituent level. The $x_1$ and $x_2$ are fractions of the incoming
momenta $P_1$ and $P_2$  of  the colliding objects with the masses $M_1$
and $M_2$. They determine the minimum energy, which
is necessary for production of the secondary particle with
the mass $m_1$ and the four-momentum $p$.
The parameter $m_2$ is introduced to satisfy the internal
conservation laws (for baryon number, isospin, strangeness, and so on).

The equation (\ref{eq:r3}) reflects minimum recoil mass hypothesis in the
elementary subprocess.
To connect kinematic and structural
characteristics of the interaction, the quantity
$\Omega$ is introduced. It is chosen in the form
\begin{equation}
\Omega(x_1,x_2) = m(1-x_{1})^{\delta_1}(1-x_{2})^{\delta_2},
\label{eq:r5}
\end{equation}
where $m$ is a mass constant and $\delta_1$ and $\delta_2$
are factors relating to the anomalous fractal dimensions  of
the colliding objects. The fractions $x_{1}$ and
$x_{2}$  are determined  to maximize the value of $\Omega(x_1,x_2)$,
simultaneously fulfilling the condition (\ref{eq:r3})
\begin{equation}
{d\Omega(x_1,x_2)/ dx_1}|_{x_2=x_2(x_1)} = 0.
\label{eq:r6}
\end{equation}
The fractions  $x_{1}$  and $x_2$ are equal to unity along
the phase space limit and
cover the full phase space accessible at any energy.

Self-similarity is a scale-invariant property connected with dropping of
certain dimensional quantities out of physical picture of the interactions.
It means that dimensionless quantities for the description of physical processes
are  used. The scaling function
$\psi(z)$ depends in a self-similar manner on the single dimensionless variable
$z$. It is expressed  via the invariant cross section
$Ed^3\sigma/dp^3$ as follows

\begin{equation}
\psi(z) = -{ { \pi s} \over { (dN/d\eta) \sigma_{in}} } J^{-1} E { {d^3\sigma} \over {dp^3}  }
\label{eq:r7}
\end{equation}
Here, $s$ is the center-of-mass collision energy squared, $\sigma_{in}$ is the
inelastic cross section, $J$ is the corresponding Jacobian.
The factor $J$ is the known function of the
kinematic variables, the momenta and masses of the colliding and produced particles.

The function $\psi(z)$ is normalized as follows
\begin{equation}
\int_{0}^{\infty} \psi(z) dz = 1.
\label{eq:r8}
\end{equation}
The relation allows us to interpret the function $\psi(z)$
as a probability density to produce
a particle with the corresponding value of the variable $z$.

Principle of fractality states that variables used in the
description of the process diverge in terms of the resolution.
This property is characteristic for the scaling variable
\begin{equation}
z = z_0 \Omega^{-1},
\label{eq:r9}
\end{equation}
where
\begin{equation}
z_0 = \sqrt{ \hat s_{\bot}} / (dN/d\eta).
\label{eq:r10}
\end{equation}
The variable $z$ has character of a fractal measure.
For the given production process (\ref{eq:r1}),
its finite part $z_0$ is the ratio
of the transverse energy released in the
binary collision of constituents (\ref{eq:r2})
and the average multiplicity density $dN/d\eta|_{\eta=0}$.
The divergent part
$\Omega^{-1}$ describes the resolution at which the collision of
the constituents can be singled out of this process.
The $\Omega(x_1,x_2)$ represents relative number of all initial
configurations containing the constituents which carry fractions
$x_1$ and $x_2$ of the incoming momenta.
The $\delta_1$ and $\delta_2$ are the anomalous fractal
dimensions of the colliding objects (hadrons or nuclei).
The momentum fractions $x_1$ and $x_2$ are determined in a way to
minimize the resolution $\Omega^{-1}(x_1,x_2)$ of the fractal
measure $z$ with respect to all possible sub-processes
(\ref{eq:r2}) subjected to the condition (\ref{eq:r3}).
The variable $z$ was interpreted  as a particle formation length.

  The scaling function of high-$p_T$ particle production, as  shown below,
  is described  by the power law, $\psi(z) \sim z^{-\beta} $.
  Both quantities, $\psi$ and $z$, are scale dependent.
  Therefore we consider the high energy   hadron-hadron interactions
  as interactions of fractals. In the
  asymptotic  region the internal structure of particles, interactions of their constituents and
   mechanism of real particle formation manifest self-similarity over a
   wide scale range.

%\end{document}

\vskip 0.5cm
{\section{Z-scaling before RHIC}}

It was established \cite{Zscal,zppg} that of data $z$-presentation
reveals the properties such as the energy and angular scaling, the power law,
$A$- and $F$-dependencies of the scaling function $\psi(z)$.
Numerous experimental data on high-$p_T$ particle spectra
obtained at U70, ISR, SpS and Tevatron used
in the analysis \cite{Zscal,zppg} are compatible each others in $z$-presentation
and give us a good reference frame for future analysis of RHIC data.

Let us remind some properties of $p_T$-presentation.
The first one is the  strong  dependence of the cross
section on energy $\sqrt s$. The second feature is a tendency
that the difference between particle yields increases
with the transverse momentum $p_T$ and the energy $\sqrt s$.
The third one is a non-exponential behavior of the spectra
at $p_{T}>4$~GeV/c.
The energy independence of data $z$-presentation
means that the scaling function $\psi(z)$ has the same shape
for different $\sqrt s$ over a wide $p_T$ range.

Figure 1(a) shows the dependence of the cross section
of $\pi^+$-meson production in $p-p$ interactions on
transverse momentum $p_T$ at $\sqrt s =11.5-53$~GeV
in a central rapidity range. The data cover
a wide transverse momentum range, $p_T = 0.2-10$~GeV/c.

Figure 1(b) demonstrates $z$-presentation of the same data sets.
One can see that the scaling function $\psi(z)$ demonstrates
independence on collision energy  $\sqrt s$
 over a wide energy and transverse momentum
range at $\theta_{cms} \simeq 90^0$.

As seen from  Figure 1(b) the scaling function  reveals
a linear $z$-dependence on the log-log scale at high-$z$.
It corresponds to the power law, $\psi(z) \sim z^{-\beta}$.
The value of the slope parameter $\beta $ is independent of
the energy $\sqrt s$  over a wide range of high transverse momentum.
This is considered as indication that
the mechanism of particle formation reveals self-similar and fractal properties.

The $p_T$-presentation demonstrates a strong angular dependence as well.
Figure 1(c) shows the dependence of the cross section of $\pi^0$-meson
production in $p-p$ collisions on transverse momentum at $\sqrt s=53$~GeV
and the center of mass angle $\theta_{cms} = (5-90)^0$.

The angular independence of data $z$-presentation means
that the scaling function $\psi(z)$ has the same shape
for different values of an angle $\theta_{cms} $ of produced
particle over a wide $p_T$  and $\sqrt s$ range.
Figure 1(b) demonstrates $z$-presentation of the
same data sets and experimental confirmation of
the angular scaling of $\psi(z)$.

The $z$-presentation of data gives indication on $F$-independence
of the scaling function \cite{Z_F}. The property means that
the scaling function $\psi(z)$
for different species of produced hadrons ($\pi^{\pm,0}, K^{\pm}, \bar p) $
at high-$z$  is described by the power law, $\psi(z) \sim z^{-\beta}$, and
the slope parameter $\beta$  is independent of flavor content of produced hadrons.
Figure 1(e) illustrates the  $F$-independence of $\psi(z)$ at high-$z$ for hadron production
in $p-Be$ collisions. For comparison of different data sets the  transformation
$
z\rightarrow (\alpha_A\alpha_F)\cdot z, \ \ \  \psi \rightarrow {(\alpha_A\alpha_F)}^{-1} \cdot \psi
$
of the scaling variable $z$ and the scaling  function $\psi$
have been used. The parameters $\alpha_A$ and $\alpha_F$ are independent
of energy $\sqrt s $ and momentum $p_T$.
The property is considered as universality of particle
formation mechanism over a wide range of small scales. We assume
that it relates to a structure of space-time itself.

\vskip 0.5cm
{\section{Z-scaling at RHIC}}

Recently the STAR and PHENIX Collaborations presented new data on inclusive
high-$p_T$ particle spectra measured  at the RHIC  in $p-p$ collisions
at $\sqrt s = 200$~GeV. In the section the data are compared  with other ones
and used as the experimental test of $z$-scaling.

{\subsection{Charged hadrons}}
The high-$p_T$ spectra of charged hadrons  produced in $p-p$ and $Au-Au$
collisions at energy $\sqrt s = 200$~GeV within $|\eta|<0.5$
were measured by the STAR Collaboration \cite{Adams}.
The results are presented in Figure 2(a).
The $p_T$-distribution of charged hadrons produced in  $Au-Au$ collisions
were measured at different centralities. The shape of the cross section drastically
changes as centrality increases.  The spectrum for $p-p$ collisions
is similar to the spectrum  observed in the peripheral $Au-Au$ collisions.
The STAR data \cite{Adams}  for $p-p$ collisions  correspond to non-single diffraction
cross section. Other experimental data correspond to inelastic
cross section. Therefore in the analysis the multiplicity particle density
$dN/d\eta$ for non-single diffraction interaction for STAR data were used.
The RHIC data and other ones for $p-p$ collisions obtained at the U70 \cite{Protvino},
 Tevatron \cite{Cronin,Jaffe} and ISR \cite{Alper}  are shown in Figure 2(b).
The charged hadron spectra were measured  over a wide  kinematic  range
$\sqrt s = 11.5-200$~GeV and $p_T = 0.5-9.5$~GeV/c.
 The strong energy dependence and the power behavior
of particle $p_T$-spectrum are found to be clearly.
The energy independence of
data $z$-presentation shown in Figure 3(c) is confirmed.
Verification of the asymptotic behavior of $\psi$ at $\sqrt s = 200$~GeV
and reach of value of $z$ up to 30 and more are of interest.

{\subsection{$\pi^0$-mesons}}
The PHENIX Collaboration published the new data \cite{Phenix} on the inclusive spectrum
of $\pi^0$-mesons produced in $p-p$ collisions in the central rapidity range
 at RHIC energy $\sqrt s = 200$~GeV.  The transverse momenta of $\pi^0$-mesons
 were measured up to 13~GeV/c.
 The $p_T$- and $z$-presentations of data  for $\pi^0$-meson spectra obtained at ISR
\cite{Angel,Kou1,Kou3,Lloyd,Eggert}
 and RHIC \cite{Phenix}  are shown in Figures 3(a) and 3(b).  One can see that $p_T$-spectra
 of $\pi^0$-meson production reveal the properties similar to that found for charged hadrons.
 The new data \cite{Phenix} on $\pi^0$-meson inclusive cross sections obtained at the RHIC
 as seen from Figure 3(b) are in a good agreement with our earlier results \cite{Zscal}.
 Thus we can conclude that the available experimental data on high-$p_T$  $\pi^0$-meson production
 in  $p-p$ collisions confirm the property of the energy independence of
 $\psi(z)$  in $z$-presentation.

{\subsection{$\eta$-mesons}}
New data on $\eta$-meson spectra in $p-p$ collisions
at $\sqrt s = 200$~GeV in the range $p_T = 1.2-8.5$~GeV/c
are presented by the PHENIX Collaboration in \cite{Hiejima}.
The $\eta/\pi^0$ ratio is found to be $0.54 \pm 0.05$ in the range
$p_T =3.5-9$~GeV/c. The value is in agreement with existing data.
We compare the data with other ones obtained at $\sqrt s = 30., 31.6, 38.8, 53.$
and 63.~GeV \cite{Kou2,E706g}.
 Data $p_T$- and $z$-presentations are shown in Figures 4(a) and 4(b).
As seen from Figure 4(b) the results of our new analysis confirm the
energy independence of the scaling function for $\eta$-meson
production in $p-p$  collisions over a wide $\sqrt s $ and $p_T$
range. Note that new result on the $\eta/\pi^0$ ratio indicates
on flavour independence of the scaling function at high-$z$.

{\subsection{$\Lambda$ and $\bar \Lambda$ hyperons}}
Here we analyze the new data obtained by   the STAR Collaboration
  \cite{Heinz} on $p_T$-spectra
of neutral strange particles ($K_S^0, \Lambda, \bar\Lambda$)
produced in $p-p$ collisions at $\sqrt s = 200$~GeV.
The transverse momentum spectra  are shown in Figure 5(a).
We have not any other data to compare with the STAR data and construct
the scaling function. Therefore the $F$-dependence of $z$-presentation
was used to determine the scaling function for $\Lambda $ and $\bar\Lambda$.
The experimental data (see Figure 3) on inclusive cross section
of $\pi^0$-mesons produced in $p-p$ collisions at $\sqrt s = 23-200$~GeV
are used to construct the asymptotics of $\psi(z)$.
The dashed line shown in Figure 5(b) is the fit of the data.
As seen from Figure 3(b) and Figure 5(b) the scaling function
is described by the power law, $\psi(z)\sim z^{-\beta}$,  on the
log-log scale at high-$z$.
The transformation of the variable $z$ and the scaling function $\psi$ for
$\Lambda $ (Fig.5(b)) and $\bar\Lambda$  in the form
$
z\rightarrow \alpha_F\cdot z, \ \ \  \psi \rightarrow {\alpha_F}^{-1} \cdot \psi
$
was used for coincidence of the asymptotics for $\Lambda, \bar\Lambda$ and $\pi^0$.
Note that the scaling function  $\psi(z)$ for $\Lambda$  reveals
different behavior at low- and  high-$z$ ranges. It is valid for $\bar \Lambda$ as well.
The parameterizations of $\psi(z)$  for $\Lambda$ and $\bar \Lambda$ were used
to predict particle spectra (see Figures 5(c) and 5(d)) at $\sqrt s =63, 200$ and 500~GeV
at high-$p_T$.

{\subsection{$\phi$-mesons}}
Recently the STAR Collaboration presented the new data \cite{Adams_phi} on spectra
of $\phi$-mesons produced  in $Au-Au$  and  non-singly-diffractive $p-p$
collisions at energy $\sqrt s = 200$~GeV.
The decay mode $\phi\rightarrow K^+K^-$ was used to reconstruct $\phi$-mesons
up to $p_T = 3.7$~GeV/c

The cross sections as a function of the difference of the transverse
mass $m_t$ and the mass of $\phi$-meson $m_{\phi}$ are shown in Figure 6(a).
The data can be used to test $z$-scaling and verify models of strange particle
formation.
The $p_T$-distribution of $\phi$-mesons  produced in $Au-Au$ collisions
measured at different centralities reveals exponential behavior.
The slope of spectra changes with the centrality.
The spectrum for $p-p$ collisions at high-$p_T$ indicates on power behavior.
The scaling functions for $\phi$ and the asymptotics for $\pi^0$ are shown in Figure 6(b).
The $F$-dependence of data $z$-presentation was used to predict $\phi$-meson spectra at
$\sqrt s =41.6, 63, 200$ and  630~GeV and $\theta_{cms}=90^0$ at high-$p_T$.
Note also that $p-p$ spectra of $\phi$-mesons is important to study the
nuclear modification factor $R_AA$ over a wide $p_T$-range
and obtain direct information about the dense nuclear matter
at hadron formation.

{\subsection{$\Xi^-$ and ${\bar \Xi}^+$ hyperons}}
The STAR Time Projection Chamber (TPC) provides excellent tracking
of charged particles with good momentum resolution \cite{NIM_TPC}.
Present statistics are sufficient to reconstruct  $\Xi^-$ and ${\bar \Xi}^+$ hyperons
over a wide $p_T$-range \cite{Witt}. The decay mode $\Xi^-\rightarrow \Lambda \pi^-$
was used to reconstruct $\Xi^-$ up to $p_T = 4.$~GeV/c.
Mid-rapidity transverse momentum spectra for $\Xi^-$ and ${\bar \Xi}^+$ from $p-p$ at
energy $\sqrt s = 200$~GeV and $|y|<0.75$ are shown in Figure 7(a).

The flavor independence of $\psi$  at high-$z$ for different pieces
allows us to construct the scaling function of $\Xi^-$ (see Fig. 7(b))
and ${\bar \Xi}^+$ over a wide $z$-range  using the asymptotics for $\pi^0$-mesons
and to predict inclusive cross sections of $\Xi^-$ and ${\bar \Xi}^+$ hyperon production
at high-$p_T$. The transverse momentum spectra for $\Xi^-$ and ${\bar \Xi}^+$ are shown
in Figure 7(c) and Figure 7(d), respectively.

As shown in \cite{Betty} the PYTHIA simulation of $\Xi^-$ spectrum
in $p-p$ collisions at $\sqrt s = 200$~GeV does not reproduce the
STAR data. Therefore our predictions can be used to tune various
PYTHIA parameters. To study azimuthal correlation of strange and
charged particles and strange tagging jet production  in $p-p$
collisions statistics should be gained. Moreover sophisticated
algorithms could essentially decrease background and increase
efficiency of strange particles reconstruction as well.

{\subsection{$K_S^0$-mesons}} Here we compare the STAR data
\cite{Heinz} for $K_S^0$-meson cross sections   with  data for
$K^+$-mesons obtained at the U70 \cite{Protvino}, Tevatron
\cite{Cronin,Jaffe} and ISR \cite{Alper} at lower energies 11.5,
19.4, 23.8, 27.4, 38.8 and 53~GeV. The data $p_T$- and
$z$-presentations are shown in Figures 8(a) and 8(b),
respectively. As seen from Figure 8(a) the energy dependence of
the cross section enhances with $p_T$. The shape of the scaling
function for $K_S^0$-mesons coincides with similar one for
$K^+$-mesons in the range $z=0.2-3.0$. It gives evidence that
mechanism of neutral and charged strange $K$-meson formation is
the same one and it reveals property of self-similarity.

{\subsection{$\pi^+$-mesons}}
The PHENIX Collaboration presented  in \cite{Harvey}
the new data on inclusive cross section of identified hadrons
($\pi^{\pm},K^{\pm}, p, {\bar p}$) produced in $p-p$ collisions
at $\sqrt s = 200$~GeV in the central rapidity range.
Transverse momentum of particles is measured up to
2.2~GeV/c. Data $p_T$- and $z$-presentations  for $\pi^+$-mesons
are shown in Figure 9.
We compare the data with another ones  obtained at the U70 \cite{Protvino},
Tevatron \cite{Cronin,Jaffe} and ISR \cite{Alper} at lower
energies $\sqrt s =11.5-53$~GeV.
As seen from Figure 9(b)  the scaling function corresponding to data \cite{Harvey}
is in good agreement with our results obtained previously \cite{Zscal}.

\vskip 0.5cm
{\section{Conclusion}}
 Analysis of new experimental data on high-$p_T$ hadrons
 ($h^{\pm},\pi^0,\eta,\Lambda,{\bar \Lambda}$, $\phi,\Xi^-,{\bar \Xi}^+$,  $K_S^0,\pi^+$)
 produced  in $p-p$ collisions  at the RHIC
 in the framework of data $z$-presentation  was performed.

The scaling function $\psi(z)$ and scaling variable $z$
are expressed via the experimental quantities, the invariant
inclusive cross section
$Ed^3\sigma/dp^3$  and the multiplicity  density of charged particles $\rho(s,\eta)$.
The scaling function $\psi$ is interpreted  as a
probability density to produce a particle with the formation length $z$.

The general regularities of high-$p_T$  particle production described
by $z$-scaling were found to be valid  in the new kinematical
range accessible at the RHIC.
Using the properties of $z$-scaling predictions of spectra
of strange particles ($ \Lambda, \bar \Lambda$, $\phi, \Xi^-, {\bar \Xi}^+$)
 produced in $p-p$ collisions at RHIC energies
in high-$p_T$ range were made. The obtained results are considered
to be useful for understanding of strangeness origin in meson and baryons.
New evidence that mechanism of particle formation reveals self-similar
and fractal properties at high-$p_T$ range  was obtained.

Thus we conclude that new data obtained at RHIC confirm the general
concept of $z$-scaling. The further inquiry and search of violation
of the scaling can give information on new physics phenomena in high energy
hadron collisions and determine domain of applicability of the strong
interaction theory.

\vskip 5mm
{\large \bf Acknowledgments.}
The author would like to thank I.Zborovsk\'{y}, Yu.Panebratsev,
O.Rogachevski and A.Kechechyan for collaboration
and numerous fruitful and stimulating discussions of the problem.

%
% *************    1(a,b,c,d,e)  *************************
\newpage

\begin{figure}
\begin{center}
\vskip -0cm
\hspace*{-7cm}
\includegraphics[width=5.2cm]{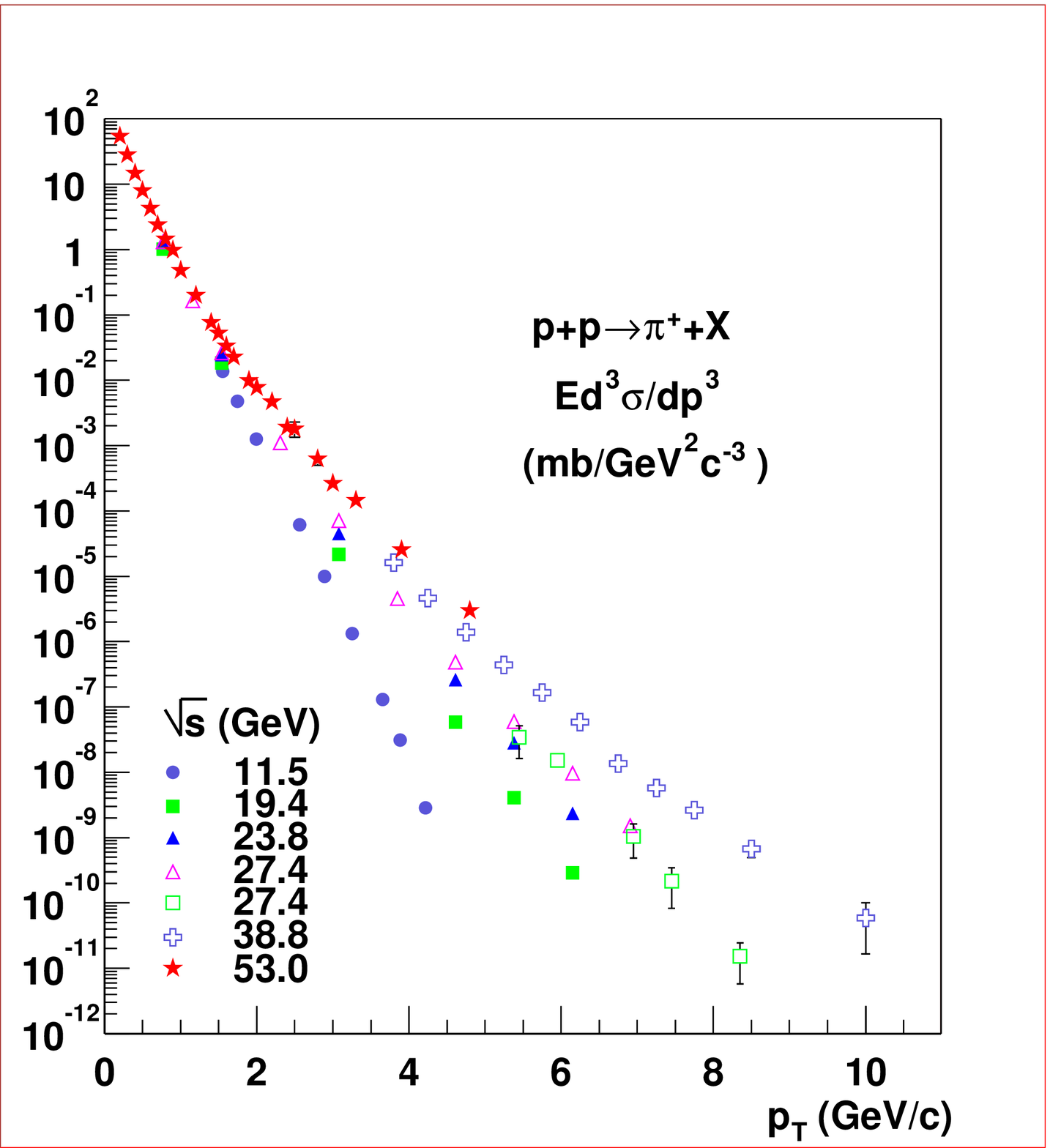}
\vskip -5.6cm
\hspace*{8cm}
\includegraphics[width=5.2cm]{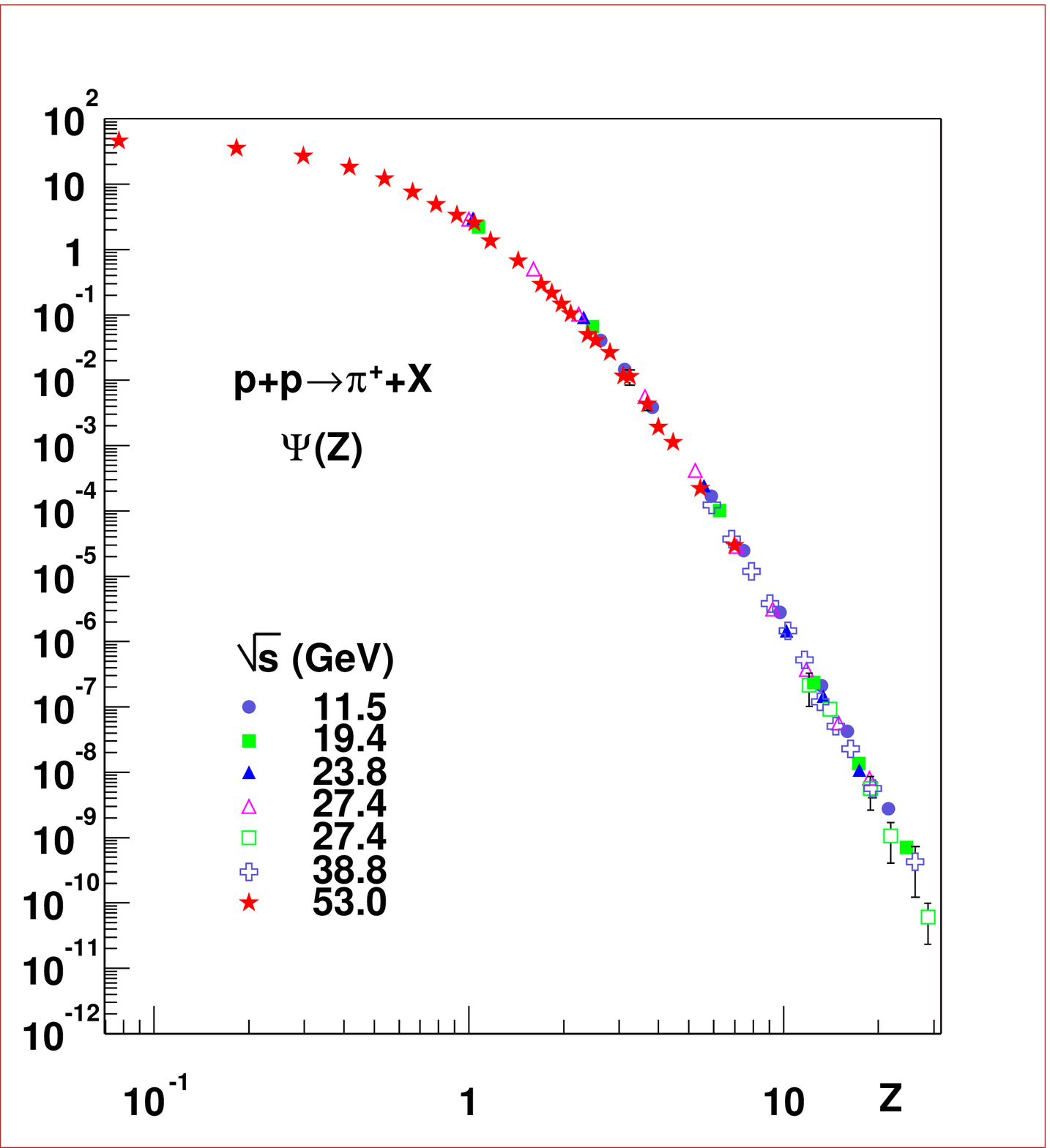}
\vskip 0.5cm

\hspace*{1cm} a) \hspace*{7cm} b)
%\end{center}
%\end{figure}

\vskip 0.5cm
\hspace*{-8cm}
\includegraphics[width=6.5cm]{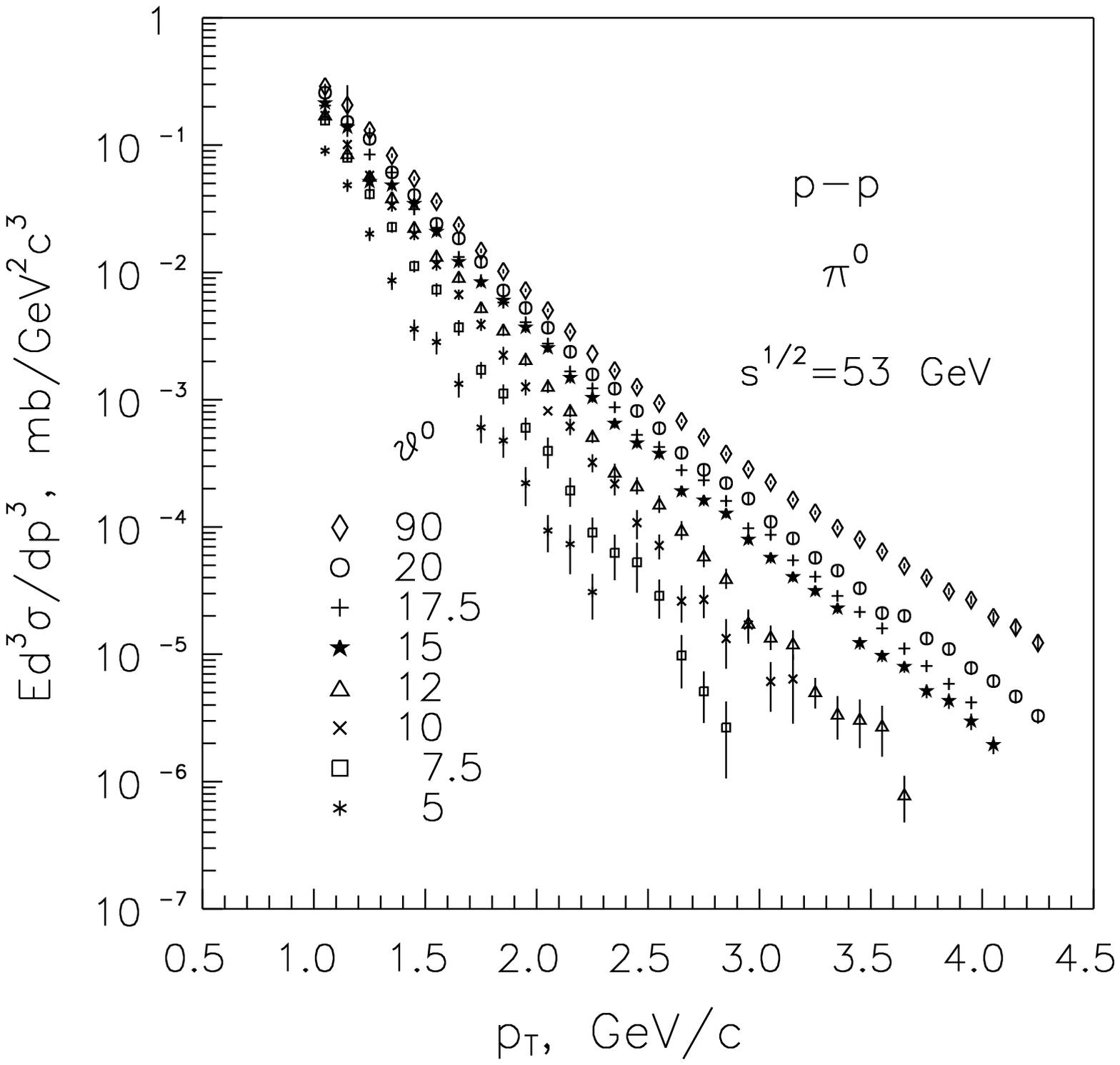}
\vskip -6.3cm
\hspace*{7cm}
\includegraphics[width=6.5cm]{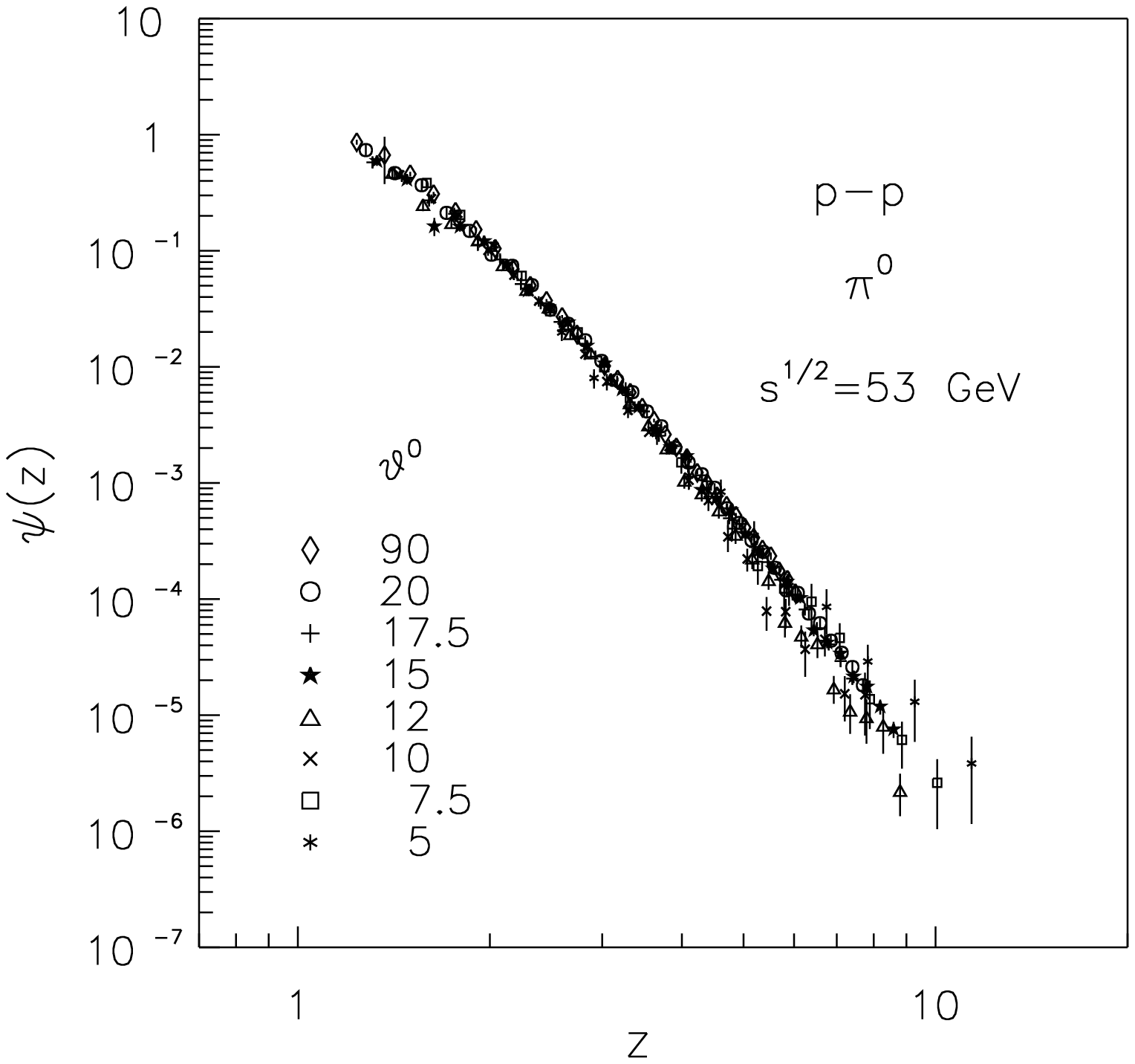}
\vskip 0.5cm

\hspace*{1cm} c) \hspace*{7cm} d)
%\end{center}
%\end{figure}

\vskip 0.5cm
%\hspace*{-4cm}
\includegraphics[width=6.5cm]{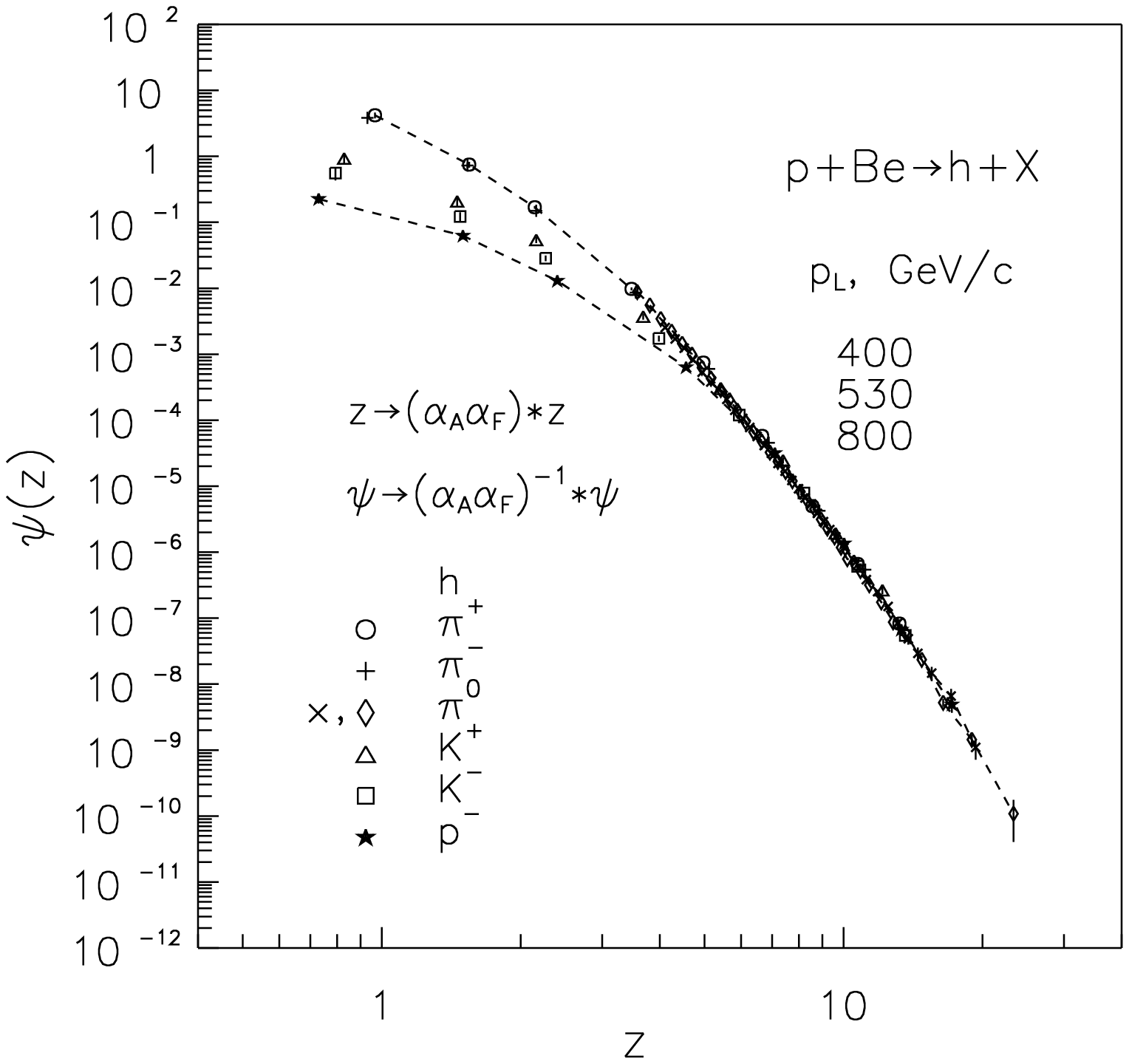}

\hspace*{1cm} e)
\end{center}

{\bf Figure 1.}
The $p_T$-(a,c) and $z$-presentations of experimental data on inclusive cross section of
particles produced in $p-p$ collisions.
The energy (b), angular (d) scaling and $F$-dependence (e) of the function $\psi(z)$.
Experimental data are taken from \cite{Adams,Protvino,Cronin,Jaffe,Alper}.
\end{figure}

%
% *************    2(a,b,c)  *************************
\newpage

\begin{figure}
\begin{center}
\vskip -3cm
%\hspace*{-7cm}
\includegraphics[width=8.2cm]{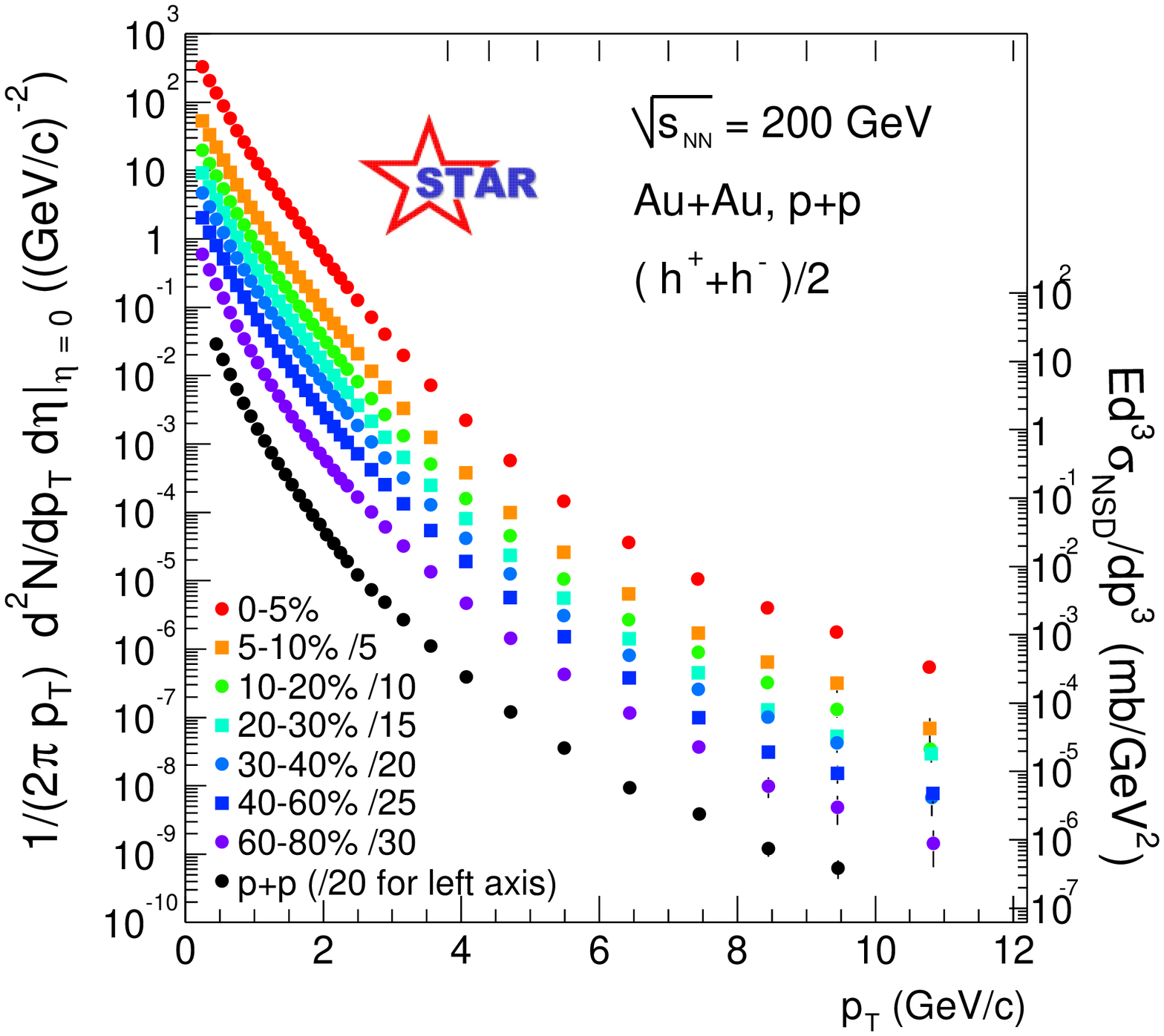}
%\vskip -5.6cm
%\hspace*{8cm}
%\includegraphics[width=5.2cm]{fig1b.eps}
\vskip 0.5cm

\hspace*{1cm} a)
%\end{center}
%\end{figure}

\vskip 1.cm
\hspace*{-8cm}
\includegraphics[width=6.5cm]{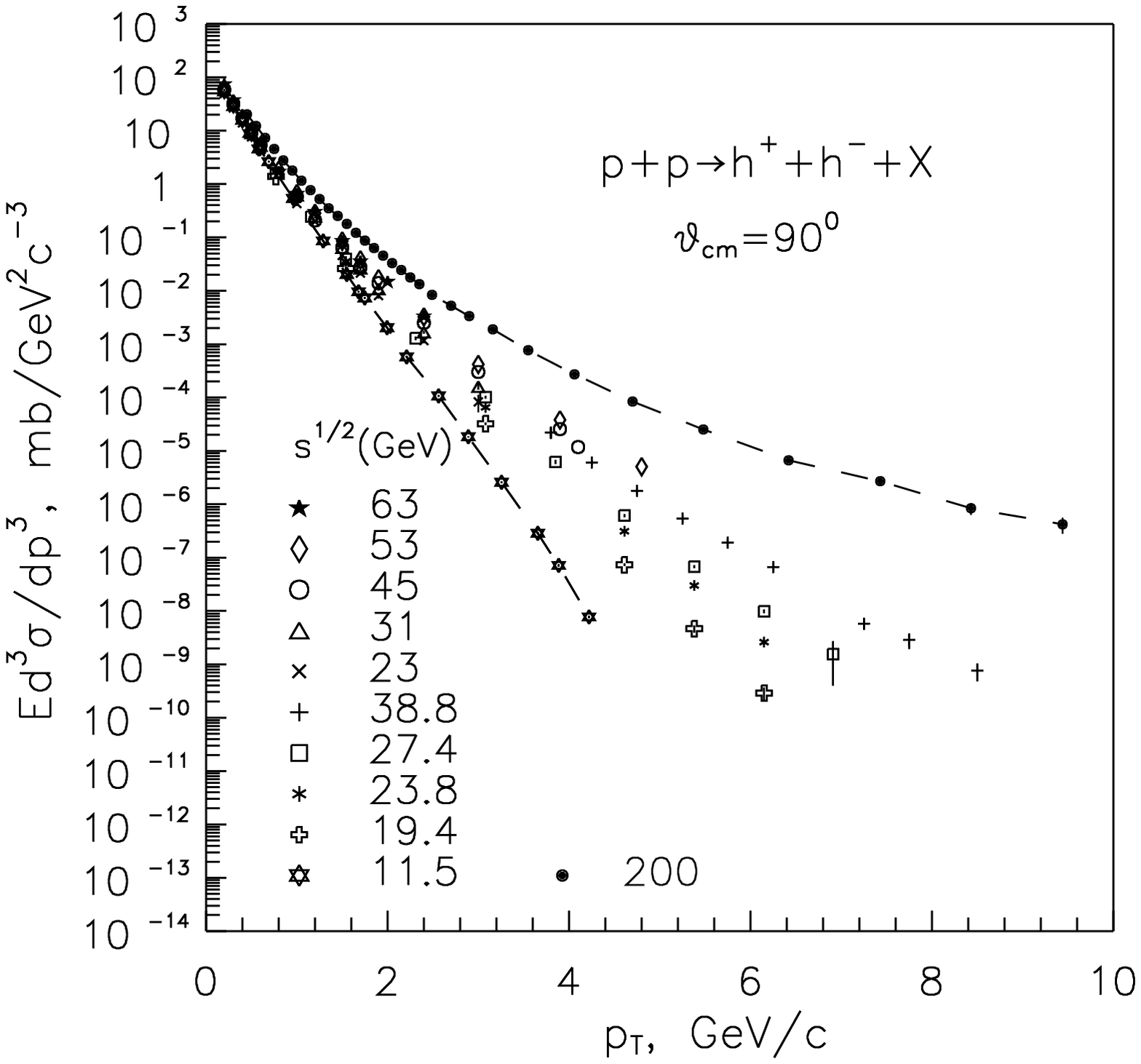}
\vskip -6.3cm
\hspace*{7cm}
\includegraphics[width=6.5cm]{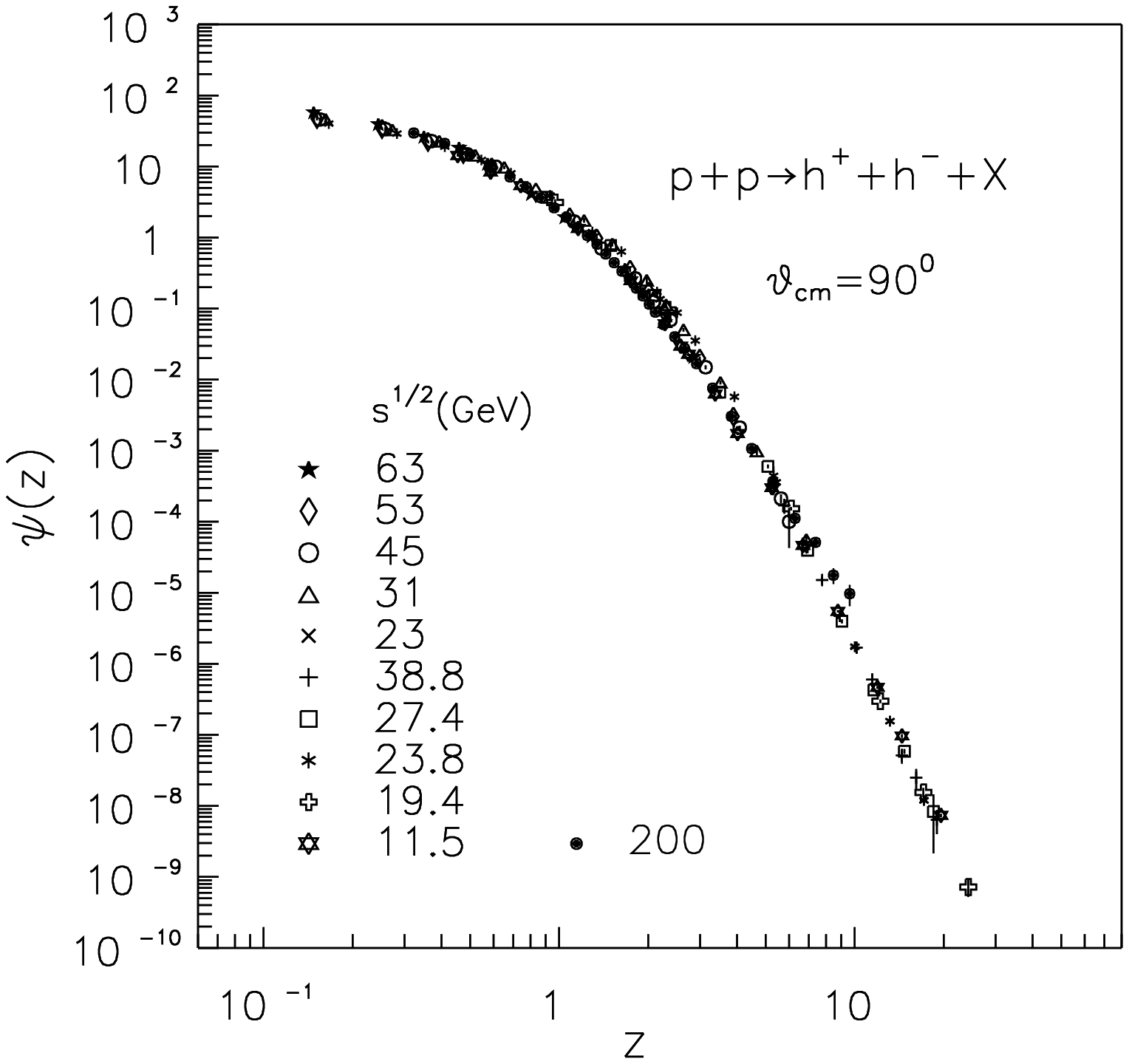}
\vskip 0.5cm

\hspace*{1cm} b) \hspace*{7cm} c)
\end{center}

{\bf Figure 2.}
 (a) Experimental data \cite{Adams} on the inclusive  cross sections
  of charged hadrons produced in $Au-Au$ and $p-p$ collisions
at $\sqrt s_{nn} =200$~GeV and  $\theta_{cms} \simeq 90^{0}$
as a functions of the transverse momentum $p_T$.
 Data (b) $p_T$- and (c) $z$-presentations of data  taken
from \cite{Protvino,Cronin,Jaffe,Alper} and \cite{Adams}.
\end{figure}

% *************    3(a,b)  4(a,b)  *************************
\newpage

\begin{figure}
\begin{center}
\vskip -3cm
\hspace*{-8cm}
\includegraphics[width=6.5cm]{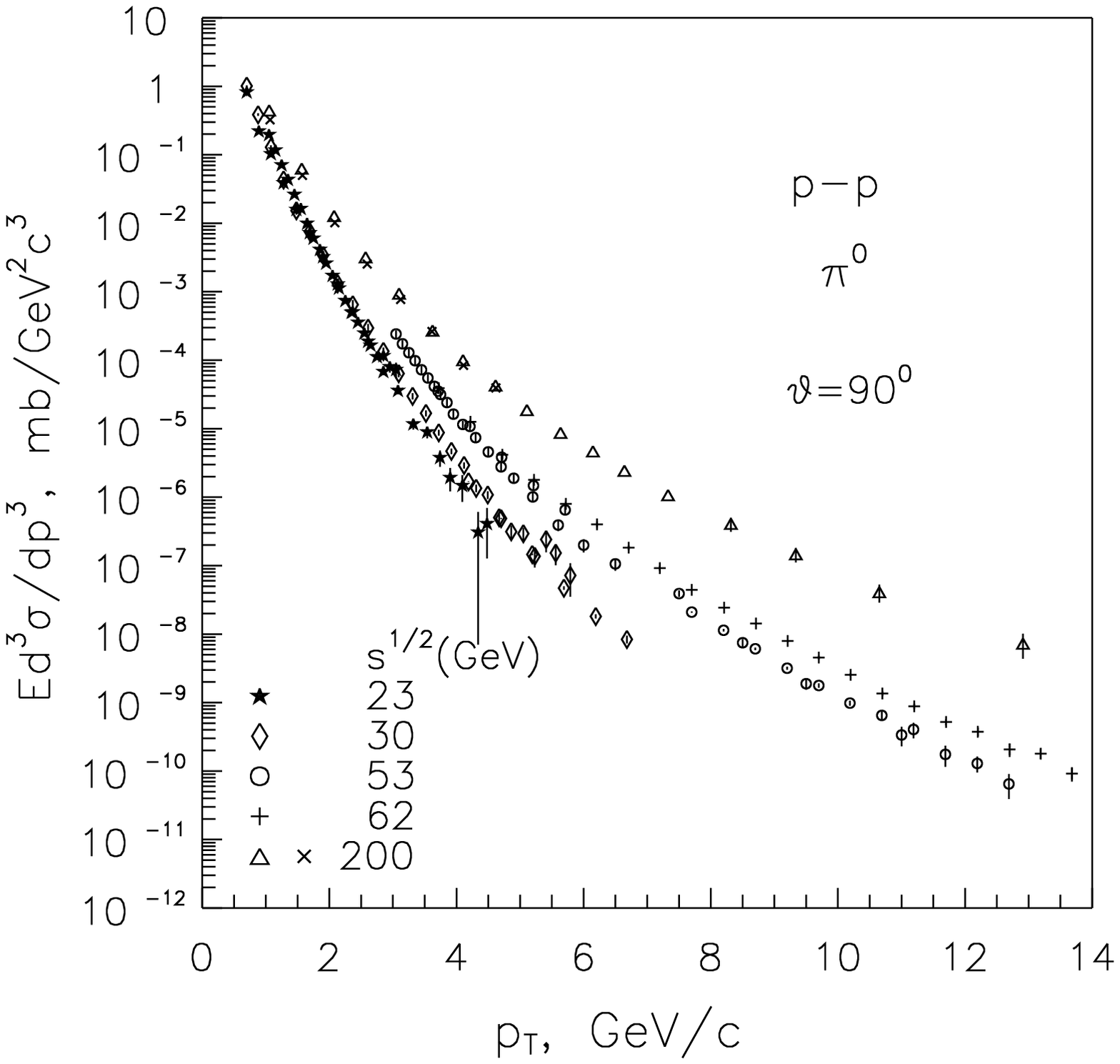}
\vskip -6.2cm
\hspace*{7cm}
\includegraphics[width=6.5cm]{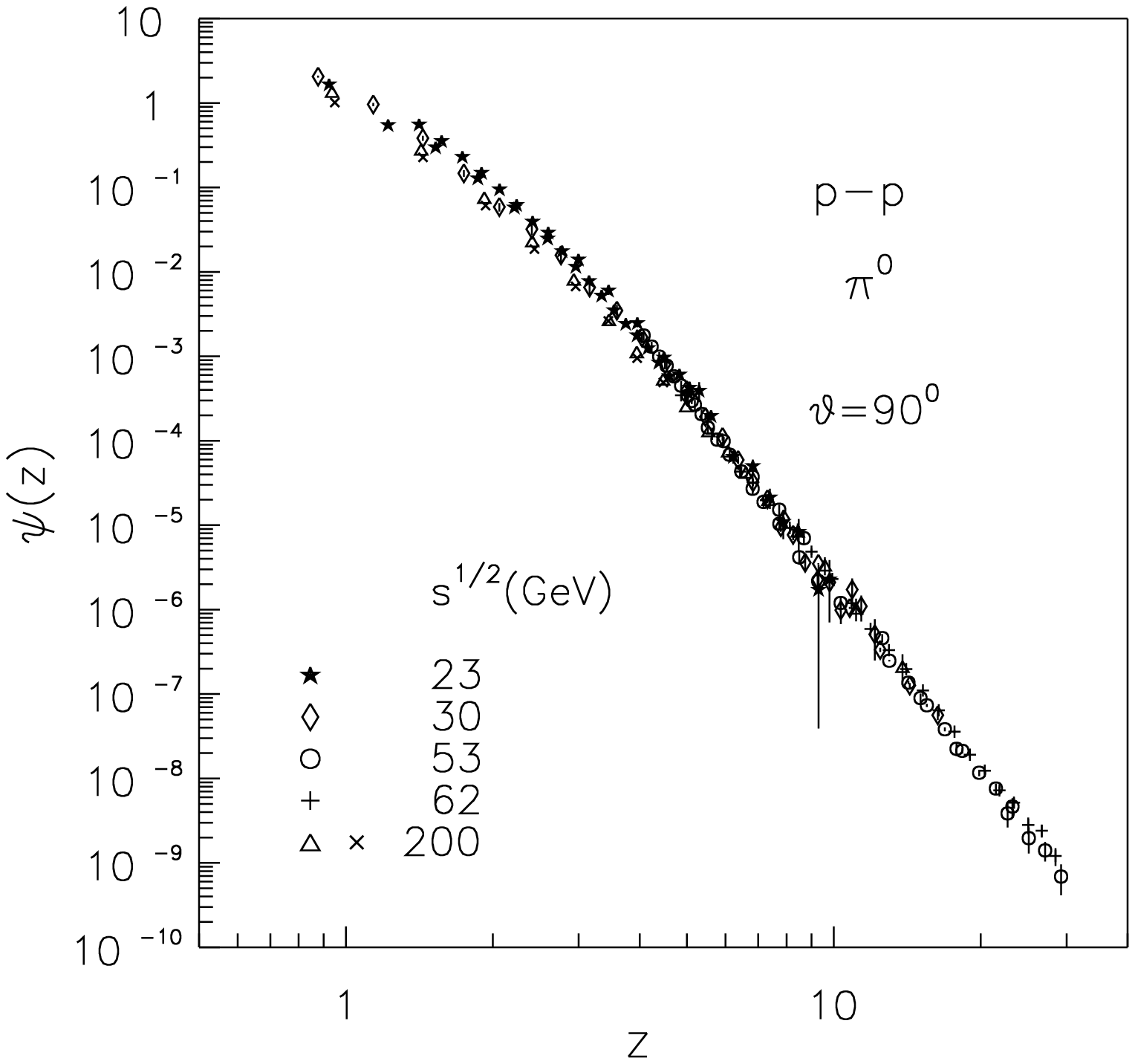}
\vskip 0.5cm
\hspace*{1cm} a) \hspace*{7cm} b)
\end{center}

{\bf Figure 3.}
The dependence of  the inclusive cross section of $\pi^0$-meson production
on the transverse
momentum $p_{T}$ in $p-p$ collisions at $\sqrt s = 30,53,62$ and 200~GeV
and the angle $\theta_{cm}$ of $90^0$.
The experimental data  are taken from
%\cite{Phenix}-\cite{Eggert}.
\cite{Angel,Kou1,Kou3,Lloyd,Eggert} and \cite{Phenix}.
(b) The corresponding scaling function $\psi(z)$.
%\end{figure}

\vskip  1.cm
%\begin{figure}
\begin{center}
\hspace*{-8cm}
\includegraphics[width=6.5cm]{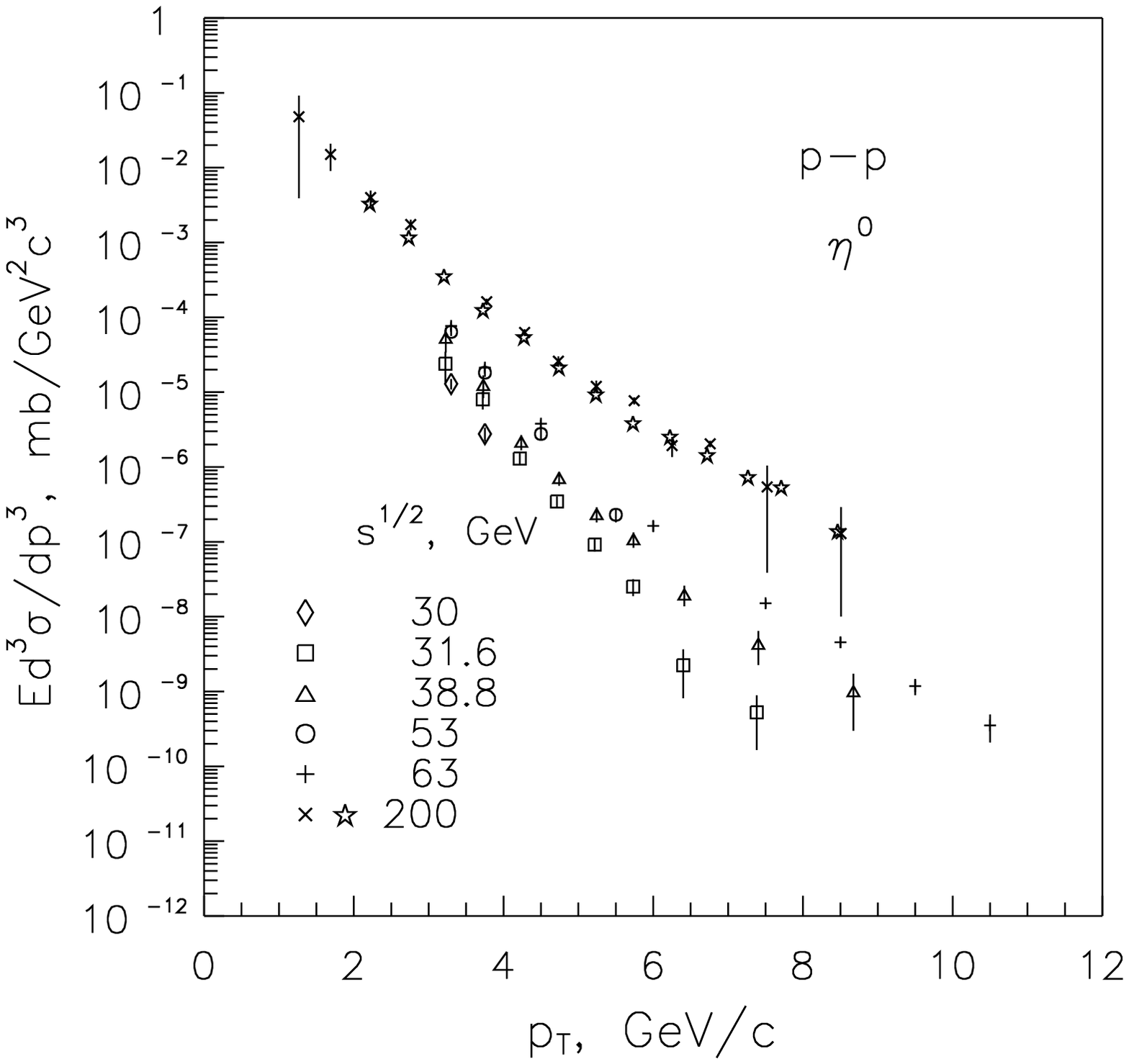}
\vskip -6.2cm
\hspace*{7cm}
\includegraphics[width=6.5cm]{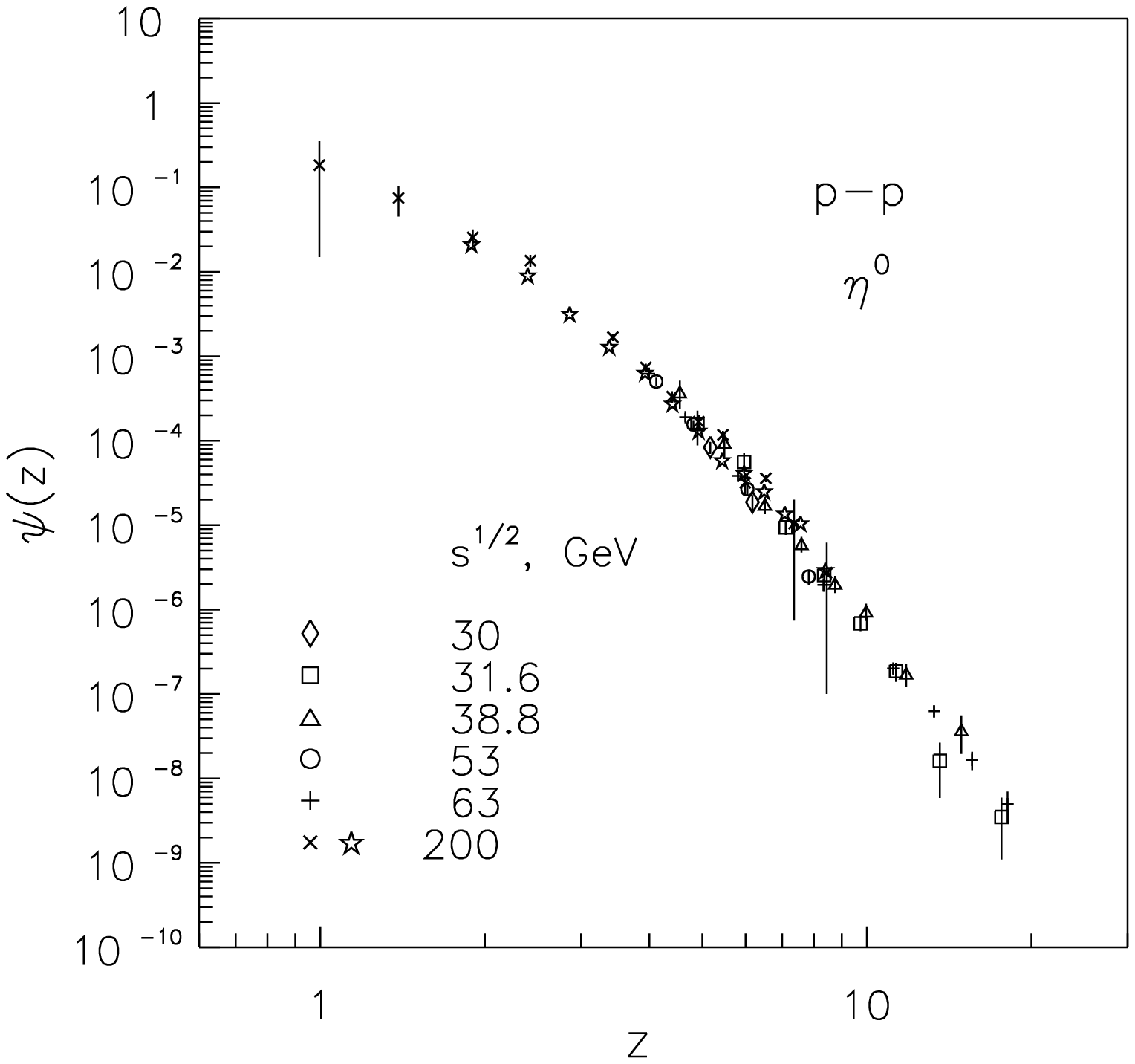}
\vskip 0.5cm

\hspace*{1cm} a) \hspace*{7cm} b)
\end{center}

{\bf Figure 4.}
 (a) The  inclusive  cross section  of
  $\eta$-mesons produced in $p-p$ collisions
in the central rapidity range as a function of the transverse momentum $p_T$
at $\sqrt s = 30-63$~GeV and 200~GeV.
Experimental data are taken from \cite{Kou2,E706g} and \cite{Hiejima}.
(b) The corresponding scaling function $\psi(z)$.
\end{figure}

% *************    5(a,b,c,d)  *************************
\newpage
\begin{figure}
\begin{center}
\vskip -1cm
\hspace*{-8cm}
\includegraphics[width=6.5cm]{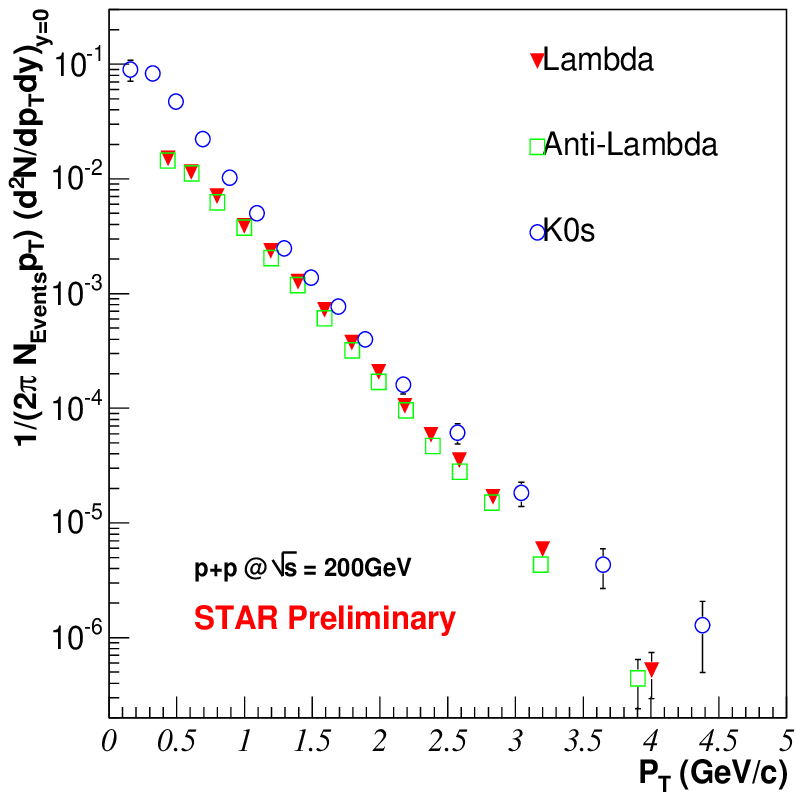}
\vskip -6.3cm
\hspace*{7cm}
\includegraphics[width=6.5cm]{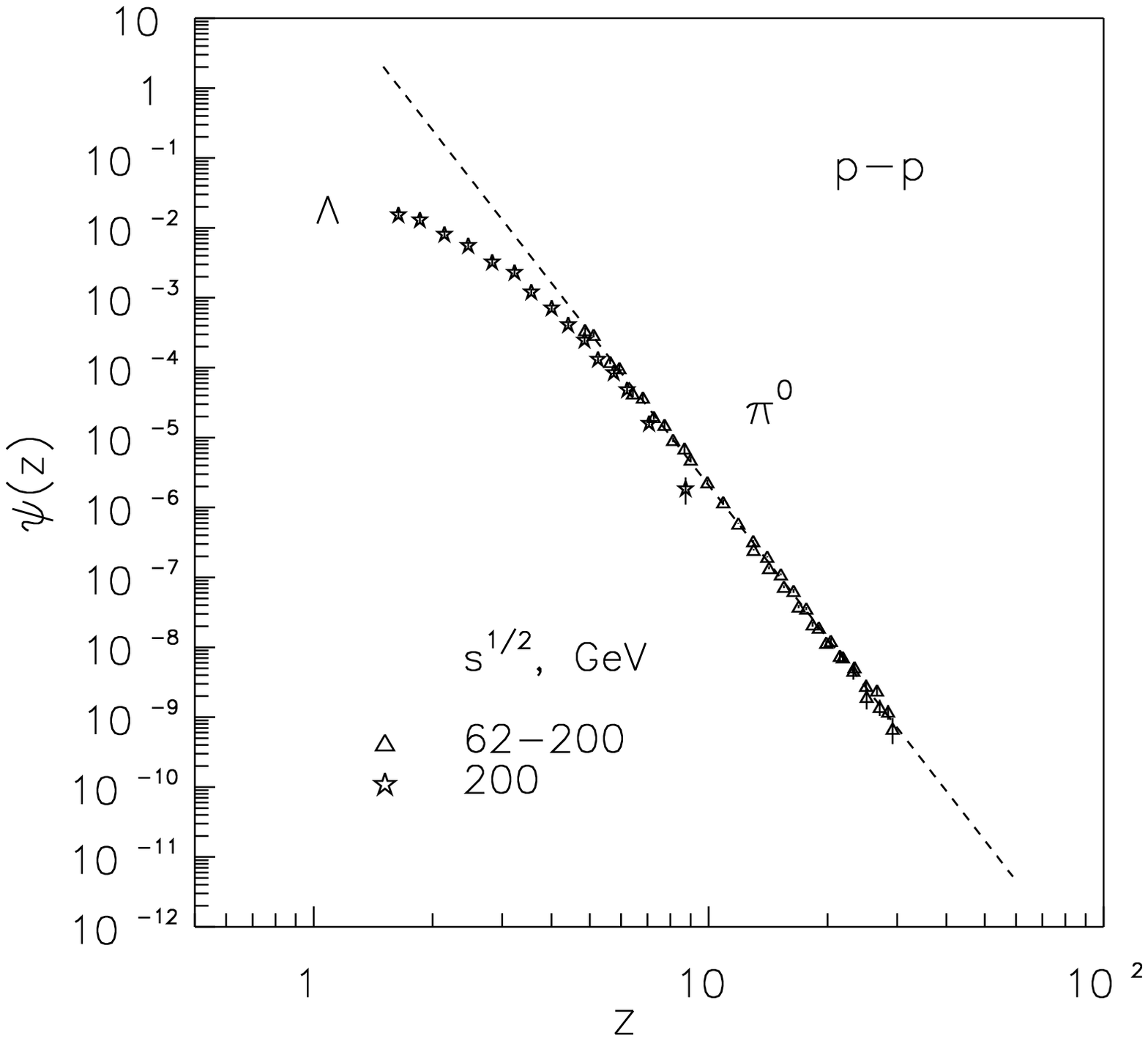}
\vskip 0.5cm

\hspace*{1cm} a) \hspace*{7cm} b)
%\end{center}
%\end{figure}

\end{center}
%\end{figure}

\vskip 1.cm

%\begin{figure}
\begin{center}
\hspace*{-8cm}
\includegraphics[width=6.5cm]{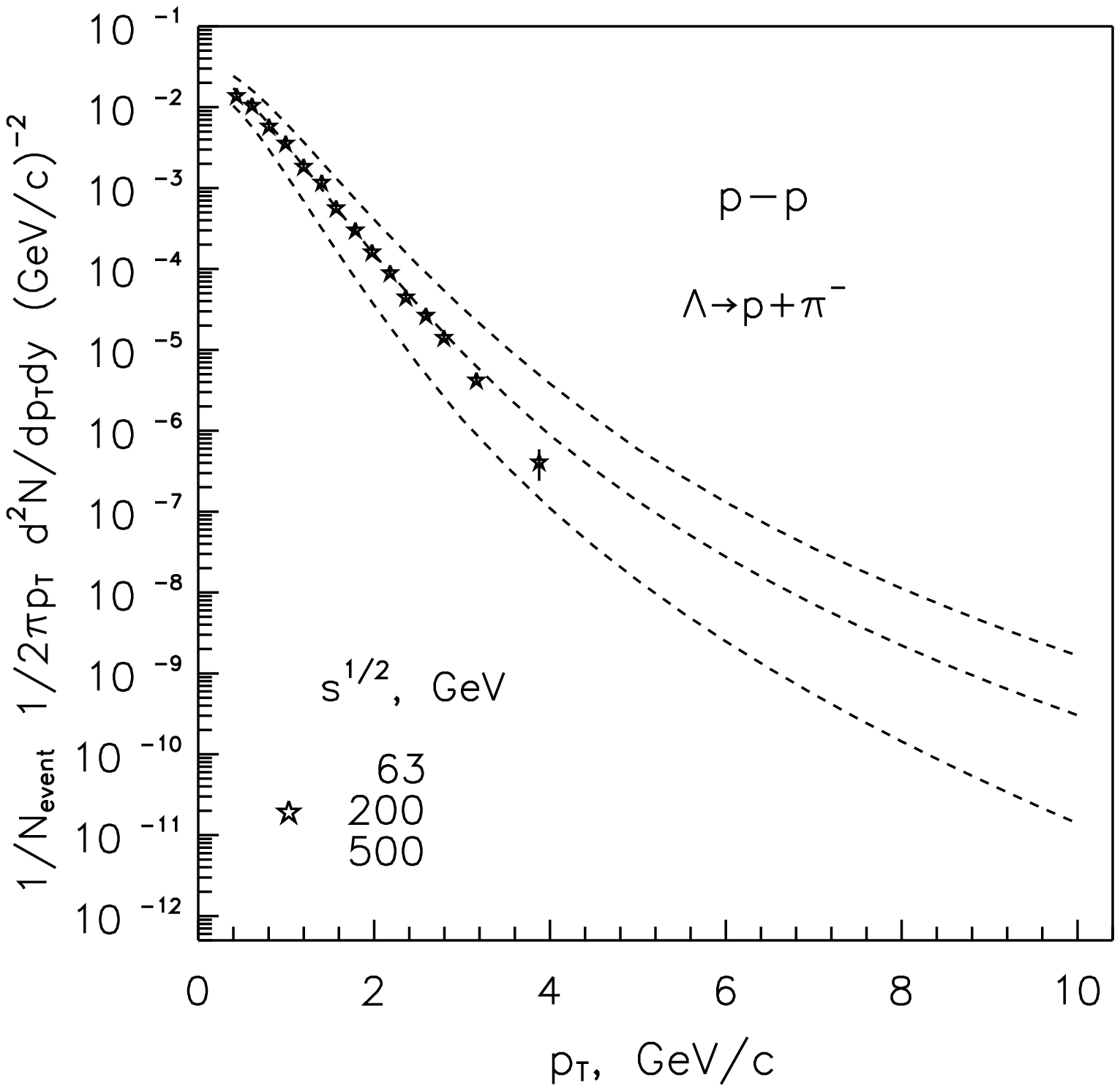}
\vskip -6.3cm \hspace*{7cm}
\includegraphics[width=6.5cm]{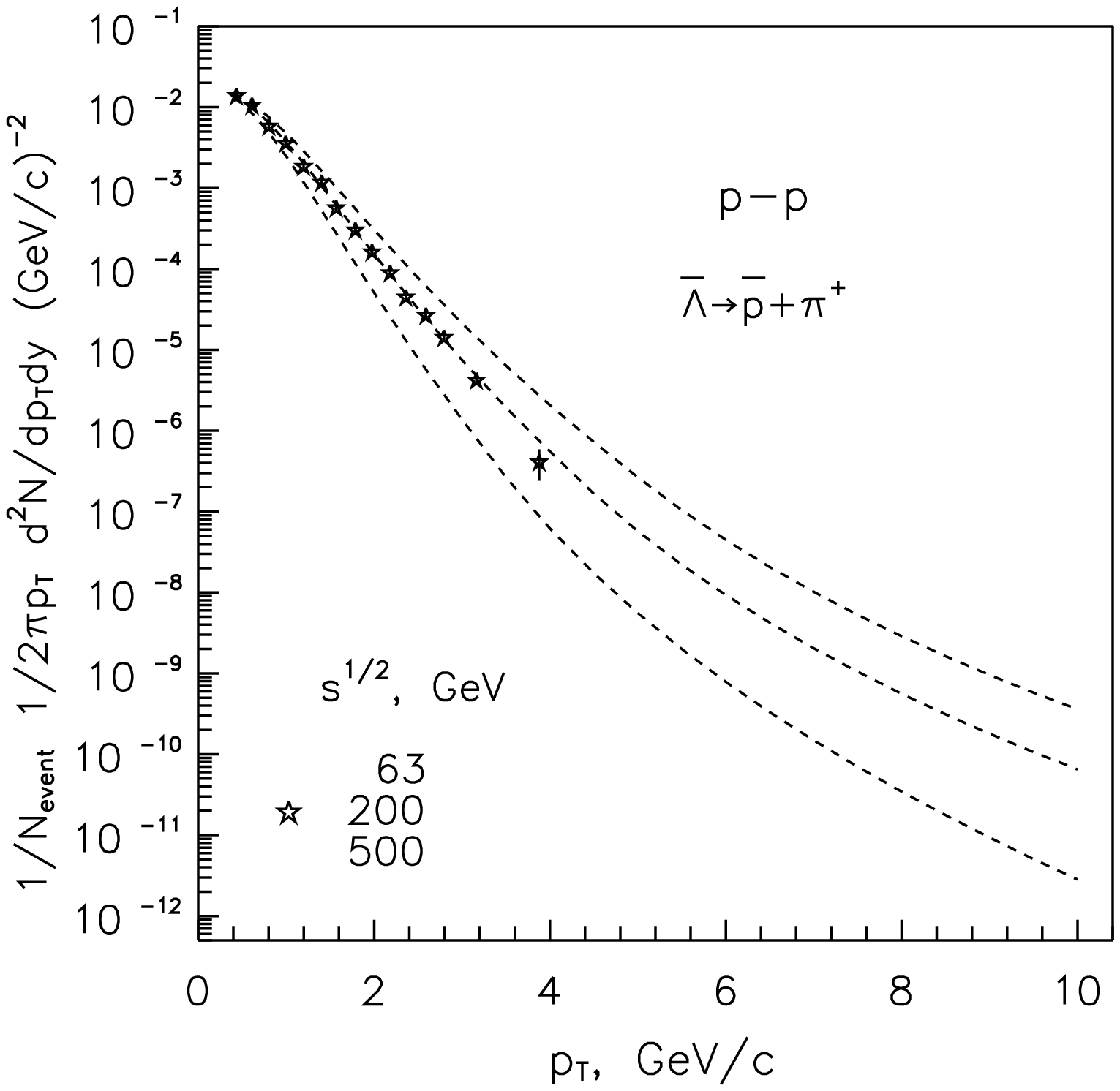}

\vskip 0.5cm

\hspace*{1cm} c) \hspace*{7cm} d)
\end{center}
{\bf Figure 5.}
(a) The inclusive  cross sections of hadrons $K_S^0,\Lambda, \bar\Lambda$ produced
in $p-p$ collisions at $\sqrt s =200$~GeV in central rapidity range
as a functions of the transverse momentum $p_T$.
Experimental data are taken from \cite{Heinz}.
(b) $Z$-presentation of  $\Lambda$ and $\pi^0$ experimental data  on cross sections.
Predictions of the inclusive spectra for $\Lambda$ (c) and $\bar\Lambda$ (d)
 production in $p-p$ collisions  at $\sqrt s = 63,200,500$~GeV and
 $\theta_{cms} \simeq 90^{0}$.

\end{figure}

% *************    6(a,b,c)  *************************
\newpage
\begin{figure}
\begin{center}
\vskip -3cm
%\hspace*{-7cm}
\includegraphics[width=6.2cm]{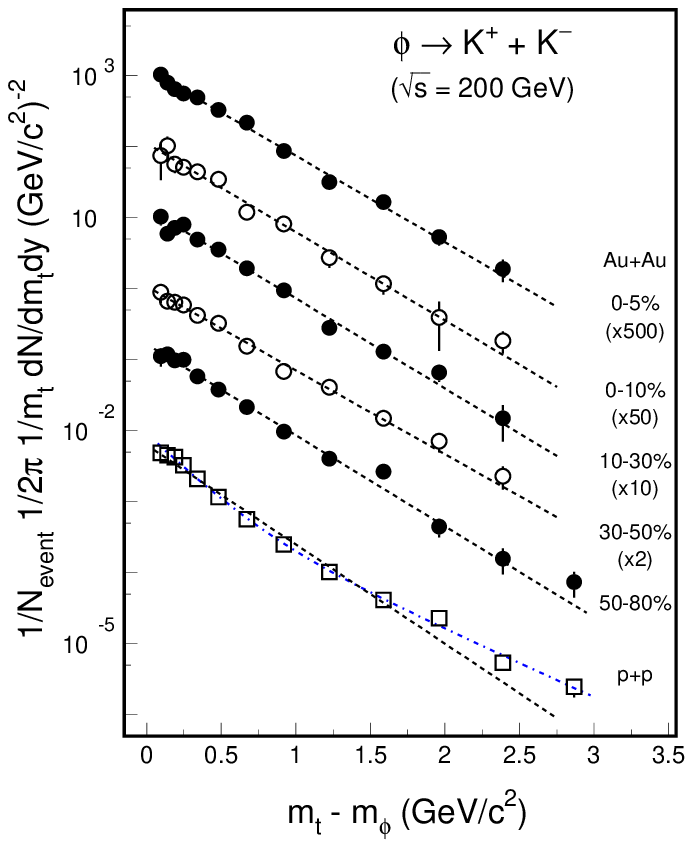}
%\vskip -5.6cm
%\hspace*{8cm}
%\includegraphics[width=5.2cm]{fig1b.eps}
\vskip 0.5cm

\hspace*{1cm} a)
%\end{center}
%\end{figure}

\vskip 1.cm
\hspace*{-8cm}
\includegraphics[width=6.5cm]{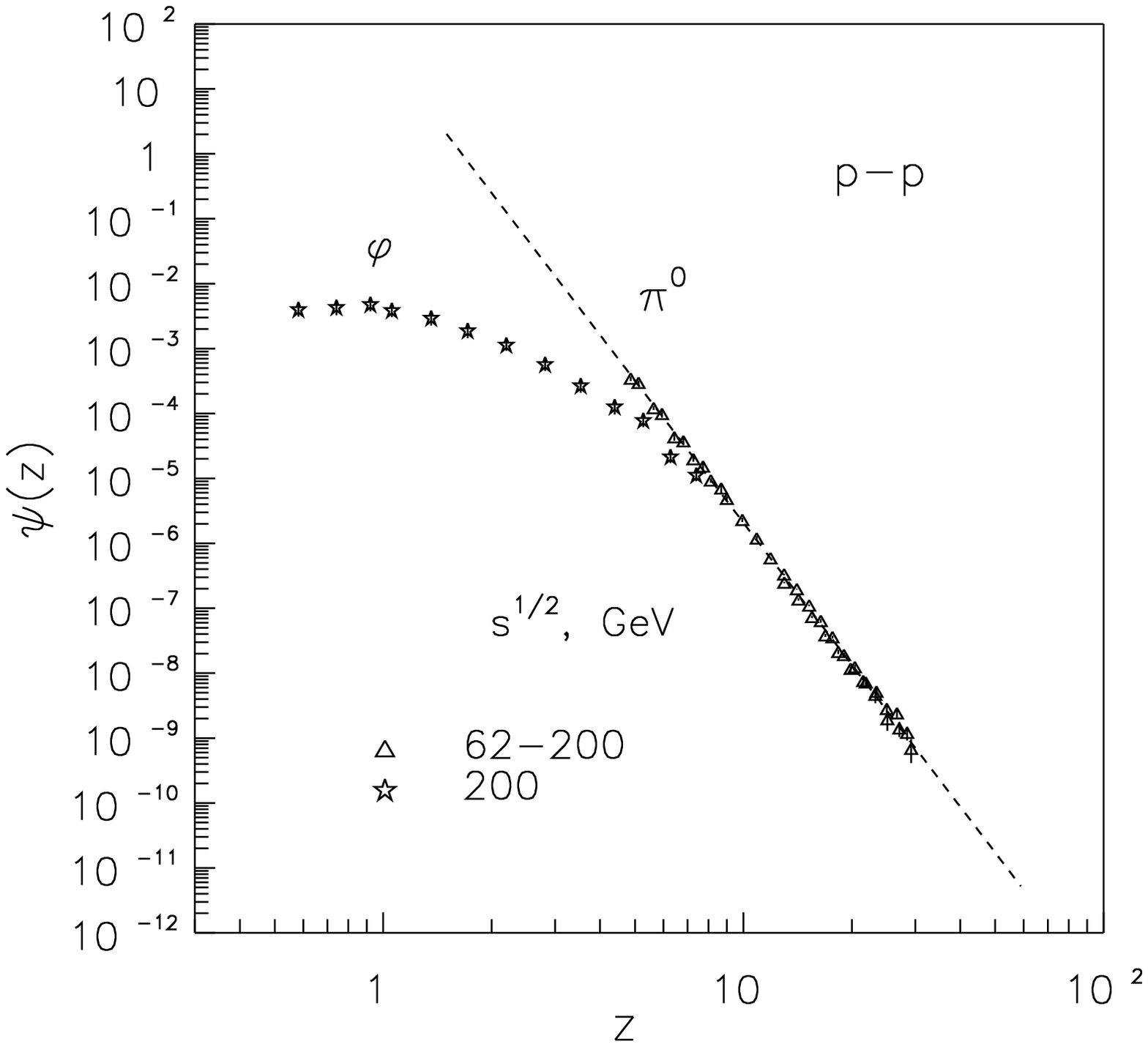}
\vskip -6.1cm
\hspace*{7cm}
\includegraphics[width=6.5cm]{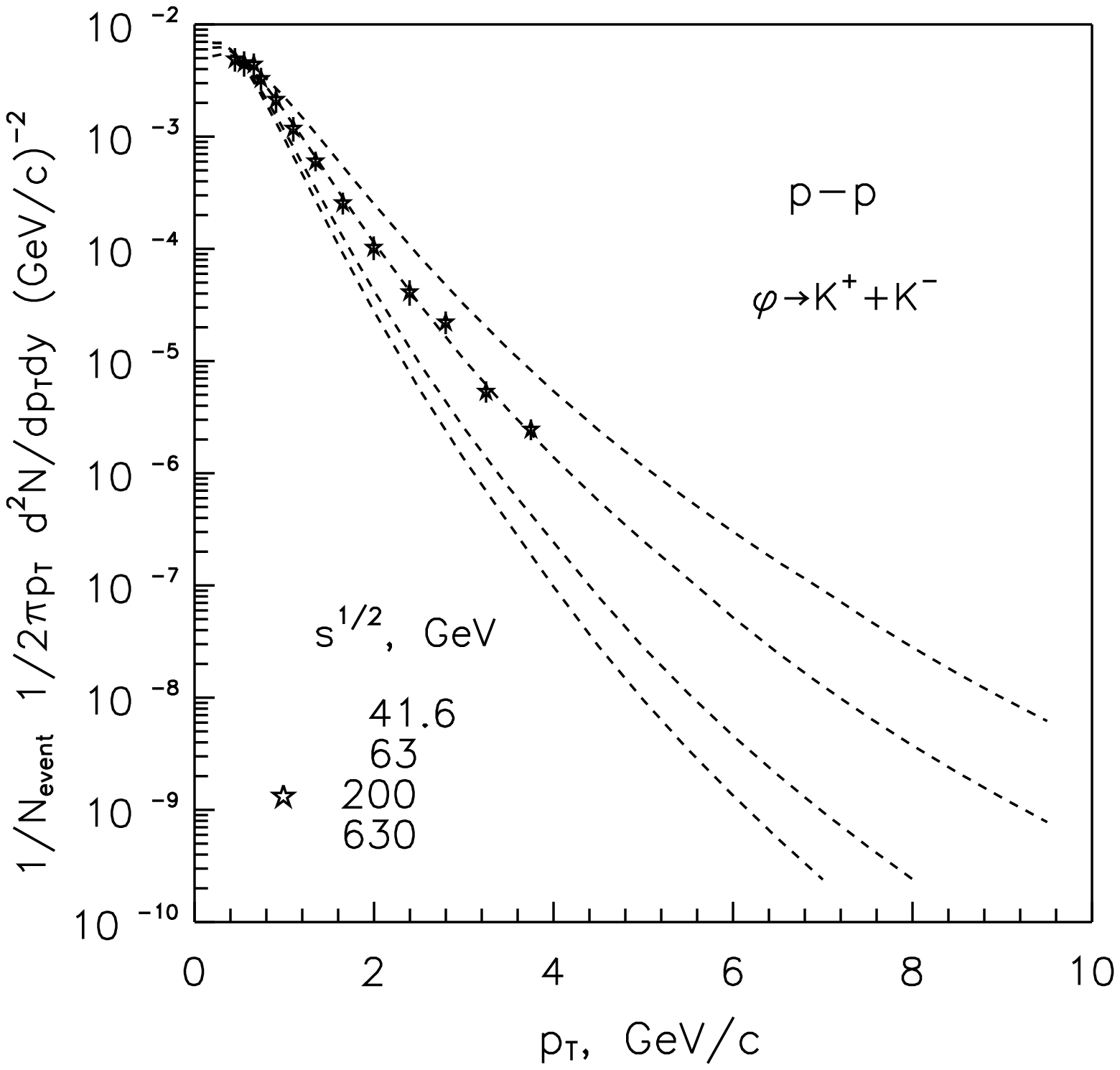}
\vskip 0.5cm

\hspace*{1cm} b) \hspace*{7cm} c)
\end{center}

{\bf Figure 6.}
(a) The inclusive cross sections of $\phi$-mesons produced
in $Au-Au$ and $p-p$ collisions at $\sqrt s_{nn} =200$~GeV in central rapidity range
as a functions of $m_t-m_{\phi}$.
Experimental data are taken from \cite{Adams_phi}.
(b) $Z$-presentation of $\phi$ and $\pi^0$ experimental data on cross sections.
(c) Predictions of the inclusive spectra for $\phi$-meson  production in $p-p$ collisions
 at $\sqrt s = 41.6,63,200,500$~GeV and  $\theta_{cms} \simeq 90^{0}$.
\end{figure}

% *************    7(a,b,c,d)  *************************
\newpage

\begin{figure}
\begin{center}
\vskip -3cm
\hspace*{-8cm}
\includegraphics[width=6.5cm]{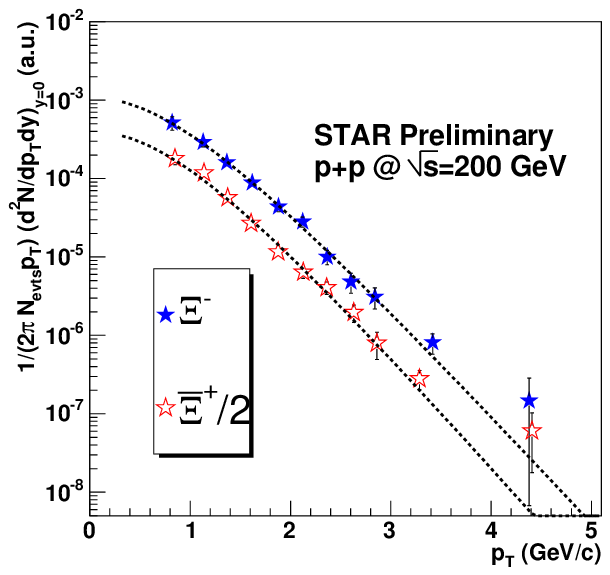}
\vskip -5.6cm
\hspace*{7cm}
\includegraphics[width=6.5cm]{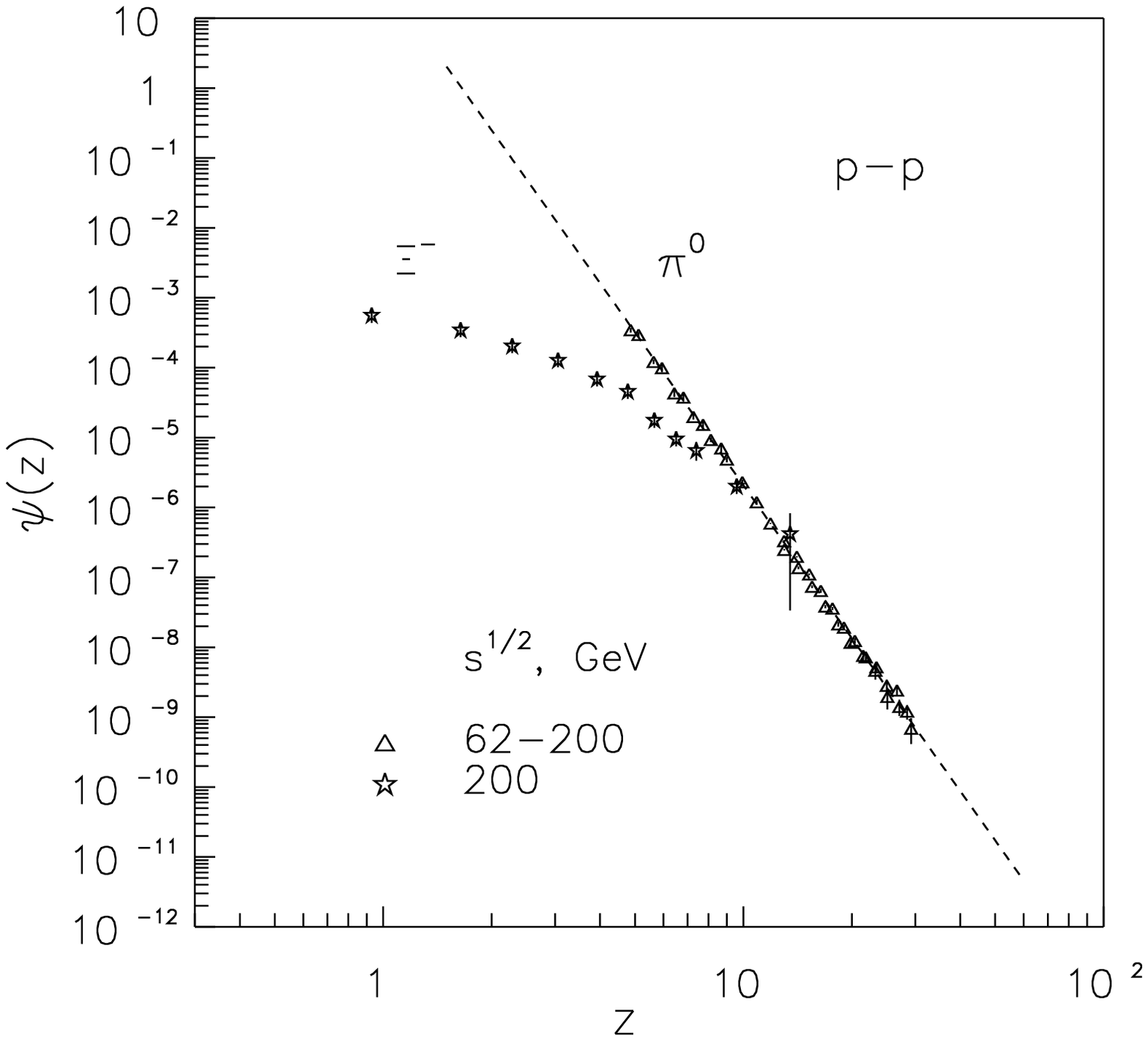}
\vskip 0.5cm

\hspace*{1cm} a) \hspace*{7cm} b)
%\end{center}
%\end{figure}

\end{center}
%\end{figure}

\vskip 0.5cm
%\begin{figure}

\begin{center}
\hspace*{-8cm}
\includegraphics[width=6.5cm]{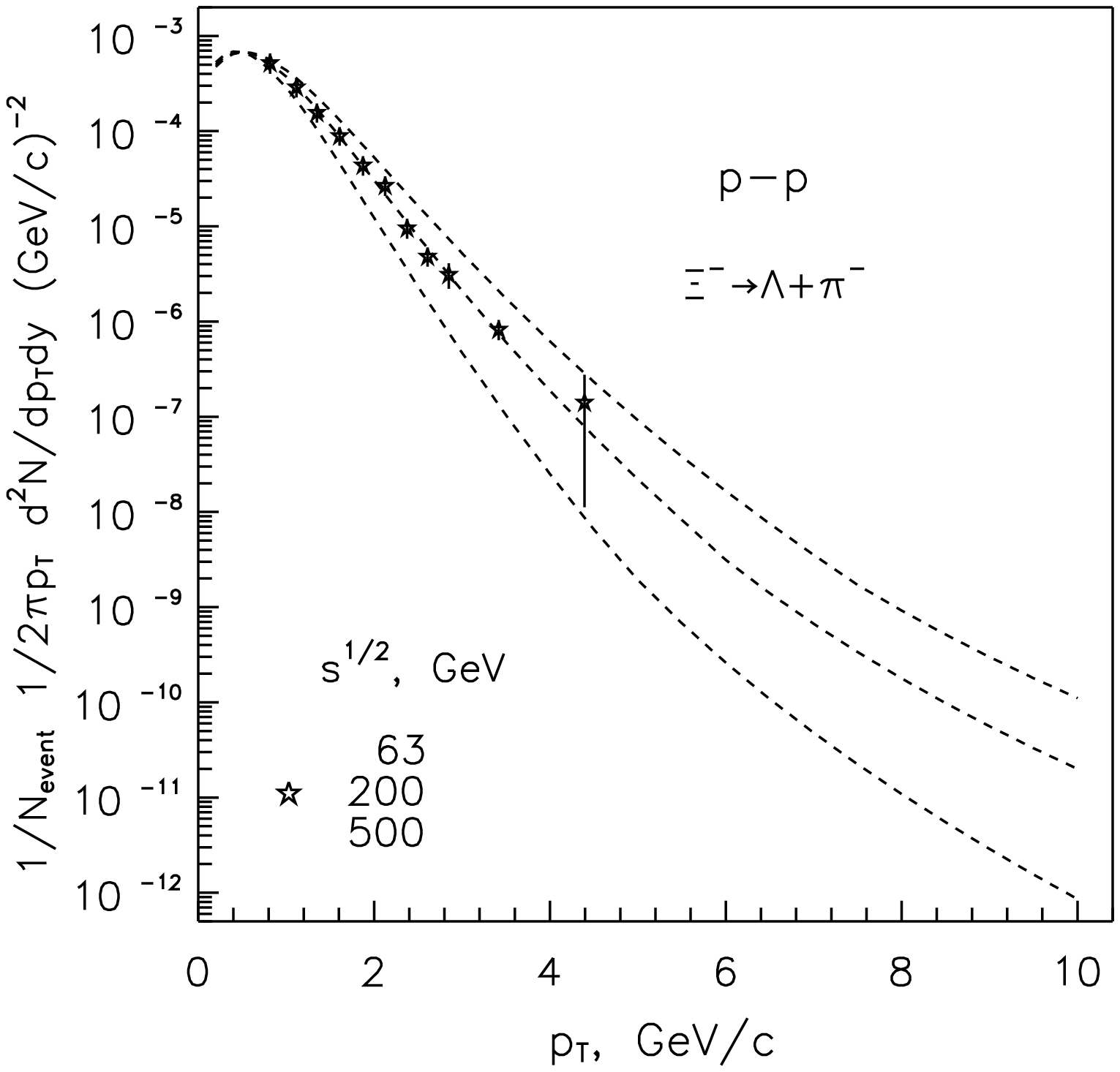}
\vskip -6.3cm \hspace*{7cm}
\includegraphics[width=6.5cm]{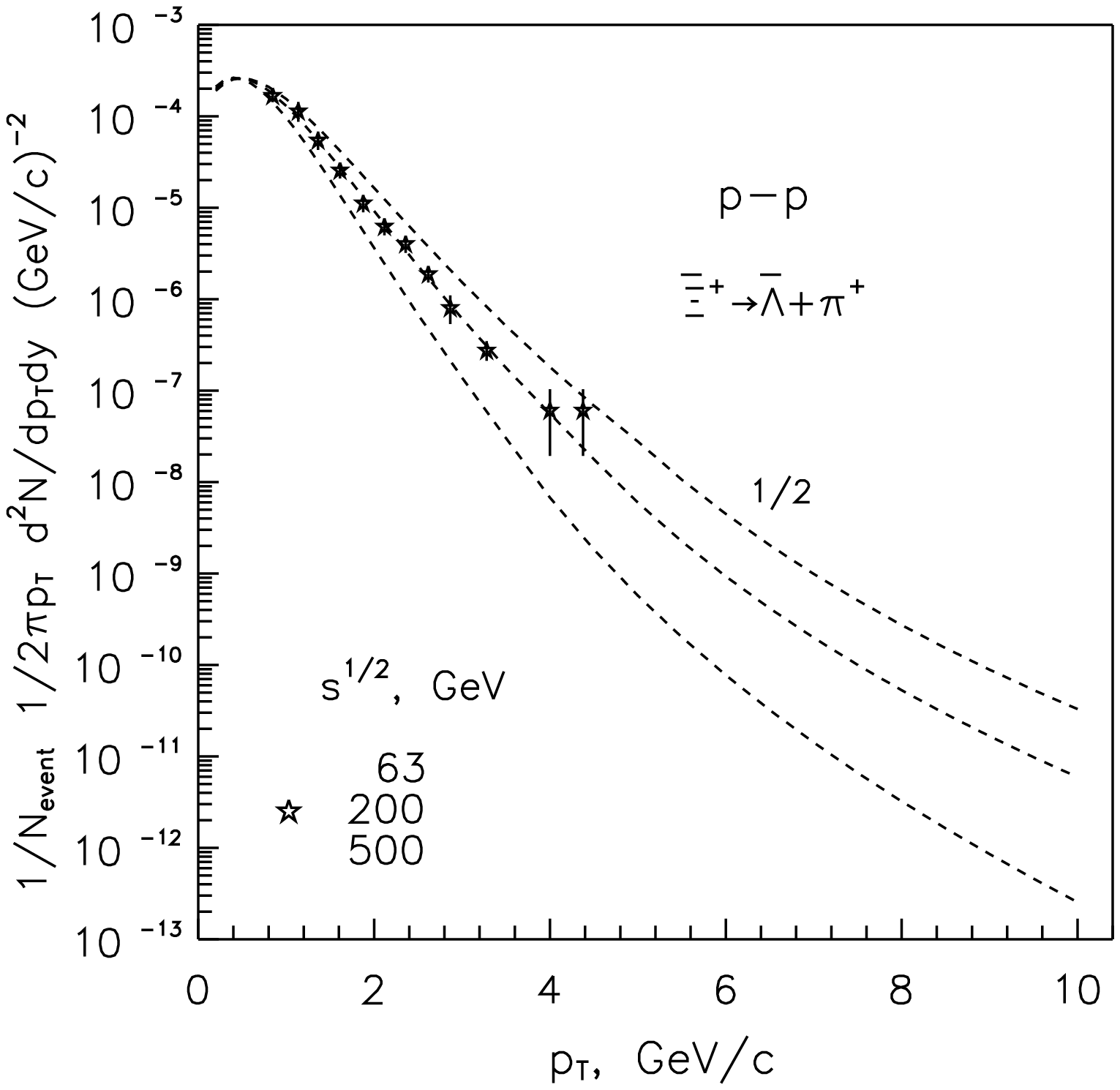}

\vskip 0.5cm

\hspace*{1cm} c) \hspace*{7cm} d)

\end{center}
{\bf Figure 7.}
(a) The inclusive  cross sections of  $\Xi^-$ and ${\bar\Xi}^+$ hyperons produced
in $p-p$ at $\sqrt s = 200$~GeV in central rapidity range
as a functions of the transverse momentum $p_T$.
Experimental data are taken from \cite{Witt}.
(b) $Z$-presentation of $\Xi^-$ and $\pi^0$ experimental data on cross sections.
Predictions of the inclusive spectra for $\Xi^-$ (c) and ${\bar\Xi}^+$ (d)
production in $p-p$ collisions  at $\sqrt s = 63,200,500$~GeV and
$\theta_{cms} \simeq 90^{0}$.
\end{figure}

% *************    8(a,b)  9(a,b)  *************************
\newpage

\begin{figure}
\begin{center}
\vskip -0cm
\hspace*{-8cm}
\includegraphics[width=6.5cm]{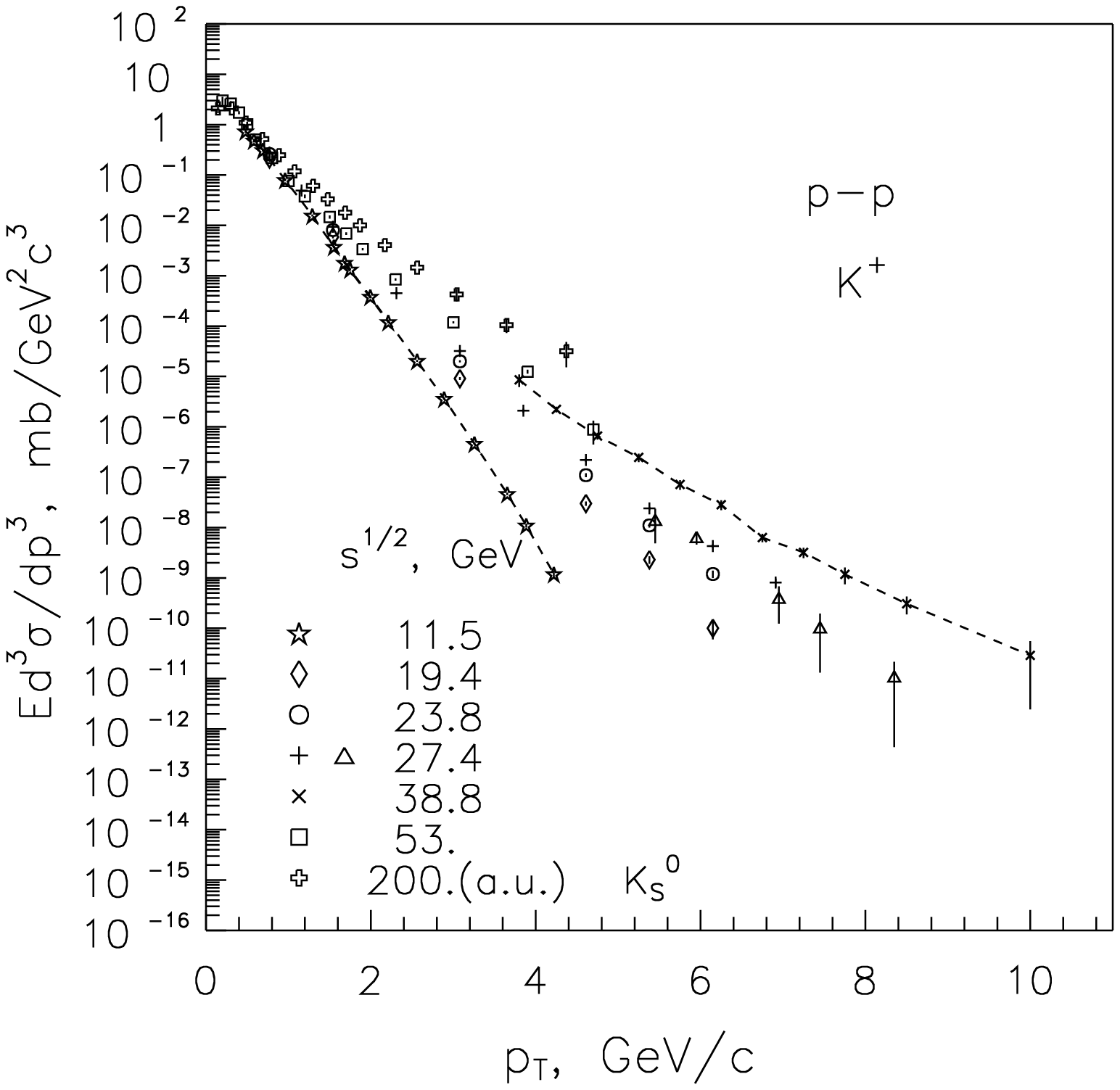}
\vskip -6.2cm
\hspace*{7cm}
\includegraphics[width=6.5cm]{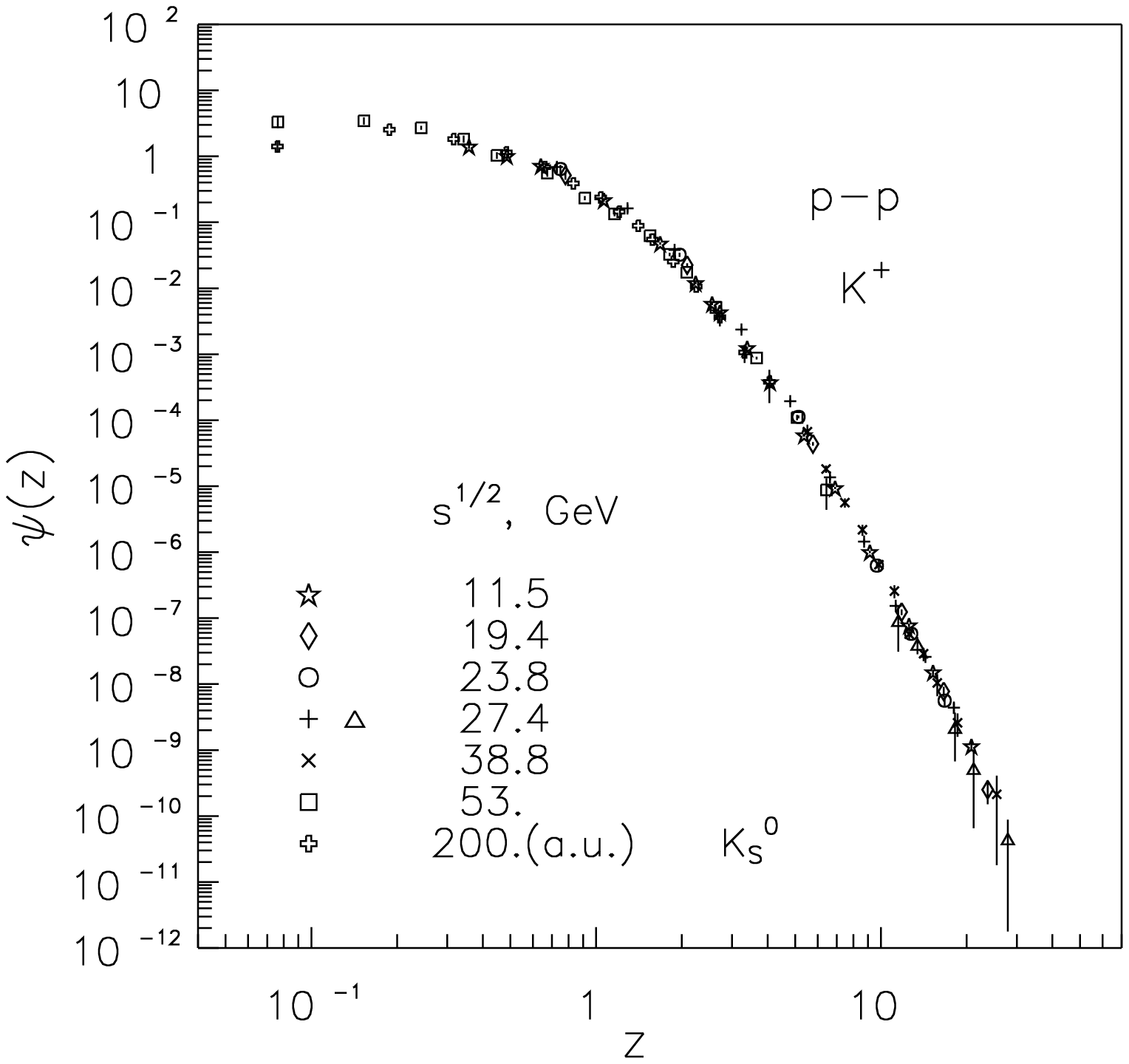}
\vskip 0.5cm
\hspace*{1cm} a) \hspace*{7cm} b)
\end{center}

{\bf Figure 8.}
(a) The inclusive  cross sections of  $K^+$- and  $K_{S}^0$-mesons
produced  in $p-p$ collisions in the central rapidity range
as a function of the transverse momentum at
$\sqrt s = 11.5-53$~GeV  and 200~GeV. Experimental data are taken
from \cite{Protvino,Cronin,Jaffe,Alper} and \cite{Heinz}.
(b) The corresponding scaling function $\psi(z)$.
%\end{figure}

\vskip 0.5cm
%\begin{figure}
\begin{center}
\hspace*{-8cm}
\includegraphics[width=6.5cm]{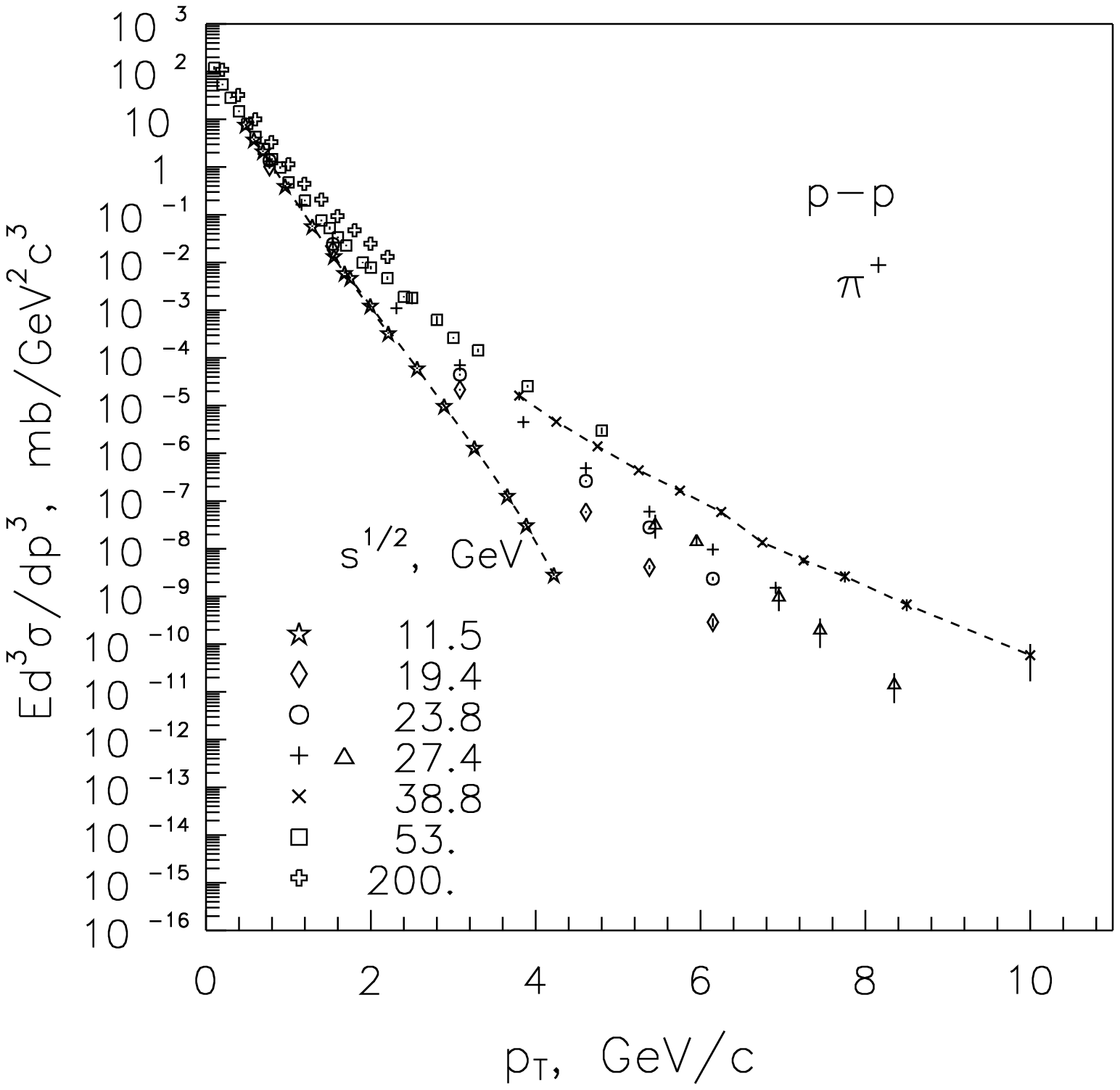}
\vskip -6.2cm
\hspace*{7cm}
\includegraphics[width=6.5cm]{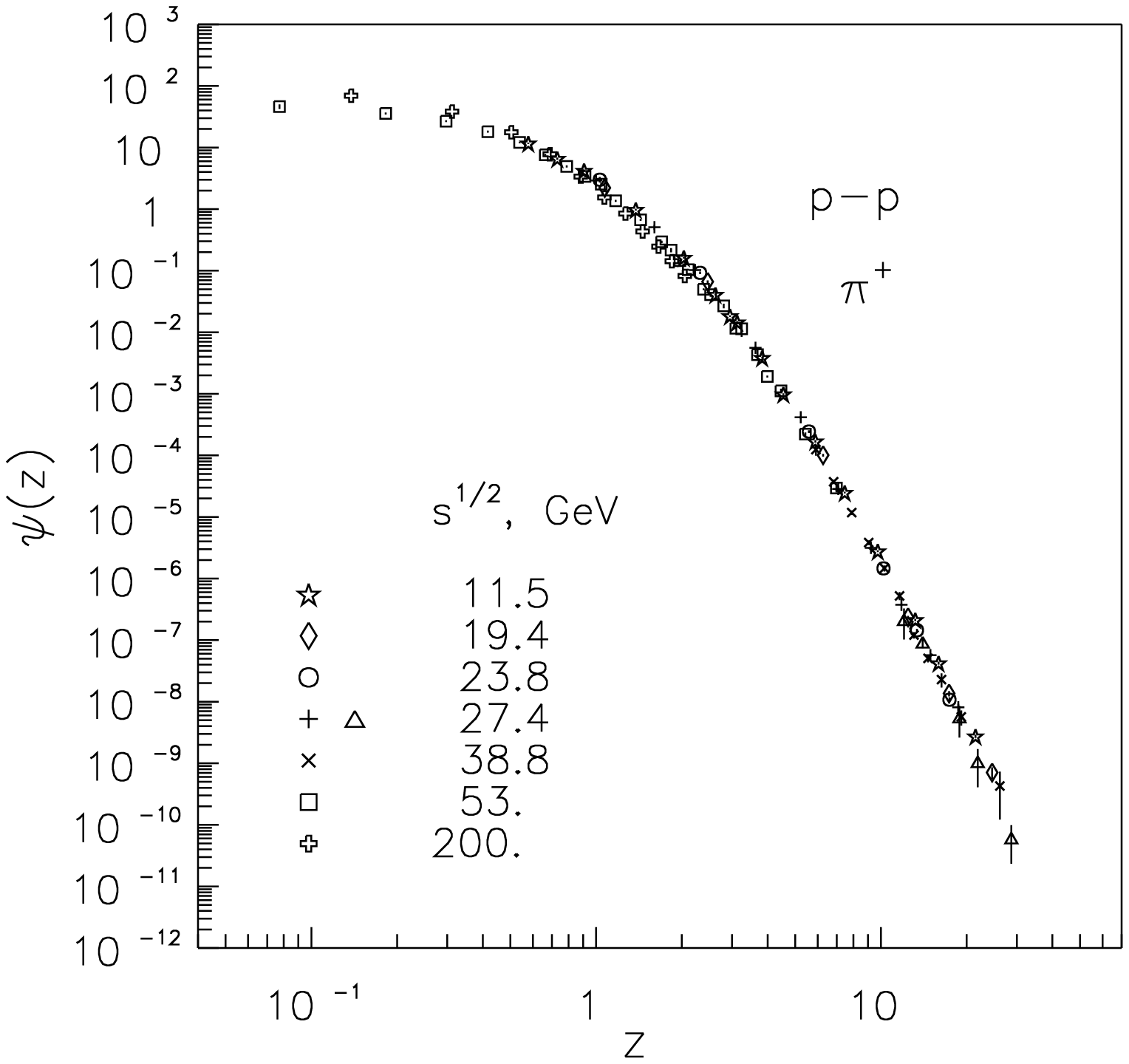}
\vskip 0.5cm

\hspace*{1cm} a) \hspace*{7cm} b)
\end{center}

{\bf Figure 9.}
(a) The  inclusive  cross section  of
 $\pi^+$-mesons produced in $p-p$ collisions
in the central rapidity range as a function of the transverse momentum $p_T$
at  $\sqrt s = 11.5-53$~GeV and  200~GeV.
Experimental data are taken from \cite{Protvino,Cronin,Jaffe,Alper} and \cite{Harvey}.
(b) The corresponding scaling function $\psi(z)$.
\end{figure}


\vskip 0.5cm
\begin{thebibliography}{99}
\bibitem{Feynman}
{\it  Feynman R.P.}//  Phys. Rev. Lett. 1969. V.23. P.1415.

\bibitem{Bjorken}
{\it Bjorken J.D.}// Phys. Rev. 1969. V.179. P.1547;\\
{\it Bjorken J.D.,  Paschanos E.A.}//  Phys. Rev. 1969. V.185. P.1975.

\bibitem{Bosted}
{\it Bosted P. et al.}// Phys. Rev. Lett. 1972. V.49. P.1380.

\bibitem{Benecke}
{\it Benecke J.  et al.}//  Phys. Rev. 1969. V.188. P.2159.

\bibitem{Baldin}
{\it Baldin A.M.}// Sov. J. Part. Nucl. 1977. V.8. P.429.

\bibitem{Stavinsky}
{\it  Stavinsky V.S.}// Sov. J. Part. Nucl. 1979. V.10 P.949.


\bibitem{Leksin}
{\it Leksin G.A.}// Report No. ITEF-147, 1976;\\
{\it Leksin G.A.}// in
{\it Proceedings of the XVIII International Conference on High
Energy Physics}, Tbilisi, Georgia, 1976, edited by N.N.
Bogolubov {\it et al.}, (JINR Report No. D1,2-10400, Tbilisi,
1977),p. A6-3.

\bibitem{KNO}
{\it Koba Z., Nielsen H.B., Olesen P.}//
Nucl. Phys. B.  1972. V.40. P.317.

\bibitem{Matveev}
{\it  Matveev V.A., Muradyan R.M., Tavkhelidze A.N.}//
  Part. Nuclei 1971. V.2. P.7.;
  Lett. Nuovo Cim. 1972. V.5. P.907.;
  Lett. Nuovo Cim. 1973.V.7. P.719.

\bibitem{Brodsky}
{\it  Brodsky S.,  Farrar G.}//
  Phys. Rev. Lett. 1973. V.31. P.1153.;
  Phys. Rev. D. 1975. V.11. P.1309.

\bibitem{Zscal}
{\it Zborovsk\'{y} I., Panebratsev Yu.A., Tokarev M.V.,\v{S}koro
G.P.}// Phys. Rev. D. 1996. V.54. P.5548.; {\it Zborovsk\'{y} I.,
Tokarev M.V., Panebratsev Yu.A.,\v{S}koro G.P.}// Phys. Rev. C.
1999. V.59. P.2227.; {\it Tokarev M.V., Dedovich T.G}// Int. J.
Mod. Phys. A. 2000. V.15. P.3495.; {\it Tokarev M.V., Rogachevski
O.V., Dedovich T.G}// J. Phys. G: Nucl. Part. Phys. 2000. V.26.
P.1671.; {\it Tokarev M.V., Rogachevski O.V.,Dedovich T.G.}// JINR
Commun. E2-2000-90. Dubna, 2000. 18p.); {\it Tokarev M.,
Zborovsk\'{y} I., Panebratsev Yu., Skoro G.}// Int. J. Mod. Phys.
A. 2001. V.16. P.1281.; {\it Tokarev M.}// hep-ph/0111202; {\it
Tokarev M., Toivonen D.}// hep-ph/0209069; {\it Skoro G.P.,
Tokarev M.V., Panebratsev Yu.A., Zborovsk\'{y} I.}//
hep-ph/0209071; {\it Tokarev M.V., Efimov G.L., Toivonen D.E.}//
 Phys. of Atom. Nucl. 2004. V.67. P.564.
{\it Tokarev M.}// Acta Physica Slovaca. 2004. V.54. P.321.

\bibitem{zppg}
{\it Tokarev M.V. }// JINR Commun. E2-98-92. Dubna, 1998. 23p.;
JINR Commun. E2-98-161.  Dubna, 1998. 10p.; {\it Tokarev M.V.,
Potrebenikova E.V}//
%JINR Preprint  E2-98-64, Dubna, 1998,
Comp. Phys.Commun. 1999. V.117. P.229.;
{\it Tokarev M., Efimov G.}// hep-ph/0209013.

\bibitem{Nottale}
{\it Nottale L.}// Fractal Space-Time and Microphysics. World Sci., Singapore, 1993.

\bibitem{Mandelbrot}
{\it Mandelbrot B.}// The Fractal Geometry of Nature. Freeman, San Francisco, 1982.

\bibitem{Imr}
{\it Zborovsk\'{y} I.}// hep-ph/0311306.

\bibitem{Z_F}
{\it Tokarev M.V.}//  In {\it Proceedings
of the International Workshop "Relativistic Nuclear Physics:
From Hundreds of MeV to TeV", Varna, Bulgaria, September 10-16, 2001,
p.280-300; JINR E1,2-2001-290.}


%Charged hadrons
\bibitem{Adams}
{\it Adams J. et al.}// (STAR Collaboration)
Phys. Rev. Lett. 2003. V.91. P.172302.

\bibitem{Protvino}
% pp
{\it Abramov V.V. et al.}//
  Yad. Fiz. 1985. V.41. P.700.;
  Pis'ma v ZhETF. 1981. V.33. P.304.;
  Yad. Fiz. 1980. V.31. P.937.


\bibitem{Cronin}
{\it Cronin J.W. et al.}// Phys. Rev. D. 1975. V.11. P.3105.;\\
{\it Antreasyan D. et al.}// Phys. Rev. D. 1979. V.19. P.764.

\bibitem{Jaffe} {\it Jaffe D. et al.}// Phys. Rev. D. 1989. V.40. P.2777.

\bibitem{Alper} {\it Alper B. et al.}// Nucl. Phys. B. 1975. V.87. P.19.

% pi+ PHENIX
\bibitem{Harvey} {\it Harvey M. et al.} // (PHENIX Collaboration)
In {\it Proceedings of the Quark Matter 2004,
 Junuary 11--17,2004,
 Oakland, California, USA};  http://qm2004.lbl.gov/.

%eta
\bibitem{Hiejima} {\it Hiejima H. et al.}// (PHENIX Collaboration)
In {\it Proceedings of the Quark Matter 2004,
 Junuary 11--17,2004,  Oakland, California, USA};  http://qm2004.lbl.gov/.

%eta in pp
\bibitem{Kou2}
{\it Kourkoumelis C. et al.}// Phys. Lett. B. 1979. V.84. P.277.

%pi0
\bibitem{Phenix} {\it Adler S.S. et al.}// (PHENIX Collaboration)
 Phys. Rev. Lett. 2003. V.91. P.241803.
%H. Torii (PHENIX Collaboration): In {\it Proceedings
%of the 16th International Conference on
%Ultra-Relativistic Nucleus-Nucleus collisions "Quark Matter 2002"},
%Nantes,   France, 18-24 July, 2002, p.753-756,
%eds. H.Gutbrod, J.Aichelin, K.Werner, Elsevier Science B.V.;
%S.S. Adler et al.: hep-ex/0304038.

\bibitem{Angel} {\it  Angelis A.L.S. et al.}//
  Phys. Lett. B. 1978. V.79. P.505.

\bibitem{Kou1} {\it Kourkoumelis C. et al.}//
  Phys. Lett. B. 1979. V.83. P.257.

\bibitem{Kou3} {\it  Kourkoumelis C. et al.}//
 Z. Phys. 1980. V.5. P.95.

\bibitem{Lloyd} {\it Lloyd Owen D. et al.}//
 Phys. Rev. Lett. 1980. V.45. P.89.

\bibitem{Eggert} {\it  Eggert K.  et al.}//
 Nucl. Phys. B. 1975. V.98. P.49.

%KOS,Lambda,barLambda
\bibitem{Heinz}   {\it Adams J., Heinz M. et al.}//(STAR Collaboration)
In {\it Proceedings of the Quark Matter 2004,
 Junuary 11--17,2004,
Oakland, California, USA}; http://qm2004.lbl.gov/; nucl-ex/0403020.

%phi
\bibitem{Adams_phi} {\it Adams  J. et al.}// (STAR Collaboration) nucl-ex/0406003.

\bibitem{NIM_TPC} {\it Anderson M. et al.}// (STAR Collaboration)
   Nucl. Instrum. Meth. A. 2003. V.499. P.659.

%Xi
\bibitem{Witt} {\it Witt R. et al.}// (STAR Collaboration) nucl-ex/0403021.

\bibitem{Betty} {\it Bezverkhny B. et al.}// (STAR Collaboration)
 Workshop "Hot Quark 2004", July 18-24, 2004, Taos Valley, New Mexico,USA.


%gamma,pi0,eta0
\bibitem{E706g}
{\it  Alverson  G. et al.}//    Phys. Rev. D. 1993. V.48. P.5.;
{\it Apanasevich L. et al.}//   Phys. Rev. Lett. 1998. V. 81. P.2642.;
%Fermilab-Pub-97-351-E;
 hep-ex/9711017;
{\it Begel M. et al.}, Workshop  "High $p_T$ Phenomena at RHIC", BNL, November 1--2, 2001.


\end{thebibliography}
\end{document}